\documentclass[showpacs,preprintnumbers,amsmath,amssymb]{revtex4}

\usepackage{graphicx}
\usepackage{dcolumn}
\usepackage{bm}

\def\mapgeq{\mathbin{\lower.3ex\hbox{$\buildrel>\over{\smash{\scriptstyle\sim}\vphantom{_x}}$}}}
\def\mapleq{\mathbin{\lower.3ex\hbox{$\buildrel<\over{\smash{\scriptstyle\sim}\vphantom{_x}}$}}}
\def\mapgeqeq{\mathbi{\lower.3ex\hbox{$\buildrel>\over{\smash{\scriptstyle\approx}\vphantom{_2}}$}}}
\def\mapleqeq{\mathbin{\lower.3ex\hbox{$\buildrel<\over{\smash{\scriptstyle\approx}\vphantom{_2}}$}}}
\def\Journal#1#2#3#4{{#1} {\bf #2} (#4) #3}

\def\NCA{Nuovo Cimento}
\def\NPB{Nucl. Phys. B}
\def\NPBOLD{Nucl. Phys.}
\def\NPSUPPL{Nucl. Phys. Proc. Suppl.}
\def\PLB{{Phys. Lett.} B}

\def\PLBOLD{Phys. Lett.}
\def\PRL{Phys. Rev. Lett.}
\def\RMP{Rev. Mod. Phys.}
\def\PRD{Phys. Rev. D}
\def\PRDOLD{Phys. Rev.}

\def\PTPSUPPL{Prog. Theor. Phys. Suppl.}
\def\PTP{Prog. Theor. Phys.}
\def\JHEP{JHEP}

\def\EPL{Europhys. Lett.}

\def\IJMP{Int. J. Mod. Phys. A}
\def\PPNP{Prog. Part. Nucl. Phys.}

\begin{document}

\preprint{TOKAI-HEP/TH-0402}

\title{
Strongly Coupled $N$=1 Supersymmetric $SO(N_c)$ Gauge Theories\\without Magnetic Degrees of Freedom
}

\author{Masaki Yasu\`{e}}
\email{yasue@keyaki.cc.u-tokai.ac.jp}
\affiliation{%
\sl Department of Physics, Tokai University,\\
1117 Kitakaname, Hiratsuka, Kanagawa 259-1291, Japan.
}%

\date{July, 2004}

\begin{abstract}
In strongly coupled supersymmetric $SO(N_c)$ gauge theories with $N_f$-quarks for $N_c-2\leq N_f(\leq 3(N_c-2)/2)$, their low-energy physics can be described by Nambu-Goldstone superfields associated with dynamical flavor symmetry breaking, which should be compared with the absence of flavor symmetry breaking in the conventional description in terms of magnetic degrees of freedom. The presence of the flavor symmetry breaking is confirmed by the well-known instanton effects in $SO(N_c)$ with $N_f=N_c-2$, which are also described by our proposed effective superpotential.  For $N_f\geq N_c-1$, our effective superpotentials utilize baryonic configurations as well as mesons composed of two quarks.  The baryonic configurations are supplied by ``diquarks" made of $N_c-1$ quarks for $N_f=N_c-1$ and by baryons composed of $N_c$ quarks for $N_c\geq N_f$.  It is argued that our effective superpotentials exhibit the holomorphic decoupling property, the anomaly-matching property and correct description of instanton effects in $SO(N_c)$ when $N_f-N_c+2$ quarks become massive if $N_f\geq N_c-1$.
\end{abstract}

\maketitle

\section{Introduction}\label{sec:1}
It is our understanding that strongly coupled $N$=1 supersymmetric (SUSY) gauge theories exhibit the $N=1$ duality \cite{Seiberg}. The $N=1$ duality is based on the physical significance of the ``magnetic" degrees of freedom such as monopoles and dyons in the low-energy description instead of the ordinary ``electric" degrees of freedom such as mesons and baryons made of ``electric" quarks.  The discovery of this $N=1$ duality has been brought about by the well-understood duality in $N=2$ SUSY gauge theories \cite{SeibergWitten}.  In order to clarify what are the magnetic degrees of freedom in the $N=1$ SUSY physics, one can perturb the corresponding $N=2$ SUSY physics \cite{BrokenN_2} to have effectively generated $N=1$ SUSY theories, where the ``electric" quarks and   gluons in $N=2$ SUSY theories are known to be continuously deformed into the ``magnetic" quarks and gluons in the corresponding $N=1$ SUSY theories.  However, it is not possible to reproduce the $N=1$ duality in all cases since $N=2$ SUSY theories are too constrained to cover all theories based on the $N=1$ duality.  For example, in supersymmetri quantum chromodynamics (SQCD) with $N_c$-colors and $N_f$-flavors, the $N=2$ duality only provides the $N$=1 duality \cite{Seiberg,EarlySeiberg} for SQCD with $3N_c/2< N_f$ ($<$ 3$N_c$), whose physics is characterized by an interacting Coulomb phase \cite{CoulombPhase}.  For SQCD with ($N_c+2\leq$) $N_f\leq 3N_c/2$, there is no direct support from the $N=2$ SQCD.  One can only assure the validity of the $N=1$ duality by listing up various indirect evidences \cite{Review} including those from brane constructions in string theory \cite{Brane}.  Furthermore, the recent analysis \cite{RecentSO} on $N=1$ SUSY $SO(N_c)$ gauge theories (with $N_f< 3(N_c-2)$) \cite{SO_IS,SO} obtained form the explicit breaking of $N=2$ SUSY theories indicates that physics near the Chebyshev point is found to be characterized by the presence of dynamical flavor symmetry breaking.  This conclusion should be contrasted with the one with the absence of flavor symmetry breaking in the $N$=1 duality although it is still possible to expect that the Chebyshev point may merge with the ordinary singular point in the $N=1$ SUSY limit and that the flavor symmetry breaking phase may disappear. 

In view of these circumstances, we have probed the possible alternative description (as the ``electric" description) \cite{Alternative,YasueSU,YasueSO} to the one based on the $N=1$ duality (as the ``magnetic" description) and have found the consistent description in terms of the Nambu-Goldstone superfields in SQCD \cite{YasueSU} and SUSY $SO(N_c)$ gauge theories (for $N_f \geq N_c$) \cite{YasueSO}.  The $N=1$ duality ensures the absence of flavor symmetry breaking while our alternative description requires the presence of flavor symmetry breaking.  These two descriptions theoretically coexist to specify possible different phases of the same gauge theories as much as the same way that QCD with two flavors theoretically allows both symmetry-preserving phase with massless proton and neutron and symmetry-breaking phase with massless pions.  Our description is at least applied to physics based on the $N=1$ duality accompanied by no direct support from the $N=2$ duality such as SQCD for $N_c+2\leq$ $N_f\leq 3N_c/2$. 

To make dynamical flavor symmetry breaking more visible in strongly coupled SUSY gauge theories crucially depends on the inclusion of a composite superfield ($S$) made of two chiral gauge superfields
\footnote{The recent discussions on $N=1$ SUSY gauge theories \cite{MatrixModel} in relation to the matrix model are based on the use of $S$.}
 introduced in Ref.\cite{VY}. It is usually neglected to analyze low-energy behavior of SUSY gauge theories because it is heavy and decoupled.  However, this decoupling is understood only after the vacuum structure is determined by effective dynamics and it is not theoretically consistent to exclude $S$ when the magnitude of its mass is unknown. Furthermore, the use of $S$ correctly describes the instanton physics in $N=1$ SUSY theories, which is one of our main concerns in this article.  Our superpotentials constructed along this line of thought in fact have provided the consistent vacuum structure with the instanton effects \cite{YasueSU,YasueSO} and have revealed the existence of the vacuum located away from the origin of a moduli space, where Nambu-Goldstone superfields show up.  The anomaly-matching conditions \cite{tHooft} are automatically satisfied by the emergence of the Nambu-Goldstone superfields \cite{AnomalyMatch}.

One may raise objections to our conclusion of the existence of the symmetry-breaking phase. These include the followings:
\begin{enumerate}
\item While there is no rigorous proof of the $N=1$ duality in some cases, its conclusion has been established by various evidences and the other conclusion such as the one in this article should be derived from a very massive number of reasonings to be convinced.
\item Because our proposed superpotential turns out to allow a symmetry preserving vacuum, the dynamics does not force nonzero vacuum expectation values (VEV's) of mesons and/or baryons. 
\item Since the origin of the moduli space is a supersymmetric minimum, then flavor symmetries are not necessarily broken.
\item Since the maximal flavor symmetry is allowed in the $N=1$ duality, the anomaly-matching is guaranteed and, if one insists that the SUSY gauge theory possesses the symmetry-breaking phase characterized by dynamical breakdown of the flavor symmetry, the anomaly-matching is automatically satisfied in any subgroups.
\item Since some of flavor-singlets such as $\det(T)$ with $T$ being flavored mesons turn out to identically vanish for $N_f>N_c$ in the classical limit, any superpotentials built out of these quantities are not so well-defined.
\end{enumerate}
These statements sound true.  However, one may carefully consider the following situations:
\begin{enumerate}
\item There are a number of established evidences but all of these are indirect in the SUSY theories with no direct proof of the validity of the $N=1$ duality and we will show the direct evidence in the $SO(N_c)$ gauge theory with $N_f=N_c-2$ that the presence of the dynamical flavor symmetry breaking is just a remnant of the instanton physics realized in the dynamics for the ``electric" quarks \cite{YasueUnpub}.
\item It is well known that any solutions determined from an appropriate superpotential yield correct SUSY vacua of the theory and there is no physical significance that picks up the symmetry preserving solution among vacuum solutions.  The inclusion of baryonic degrees of freedom 
\footnote{We use the word "baryon" to specify composites made of color-antisymmetrized quarks.}
allows us to have a solution with nonzero VEV's as in SQCD with $N_f=N_c$.
\item It is dynamically true only if the 't Hooft anomaly-matching conditions are satisfied.  If the 't Hooft anomaly-matching conditions are not satisfied at the origin of the moduli space, then the theory is unstable at the origin.  Usually, this instability is removed by the appearance of the magnetic degrees of freedom but it may also trigger the symmetry-breaking so that the flavor symmetry breaks all the way down to the point, where all the anomalies are saturated by the Nambu-Goldstone superfields associated with this symmetry breaking. There are ordinary Nambu-Goldstone bosons that saturate the anomalies of broken chiral symmetries. In the SUSY gauge theories, there are also massless fermions contained in the Nambu-Goldstone superfields, whose anomalies must match with those of residual chiral symmetries.  If all of the anomalies are balanced, then dynamical symmetry breakdown ceases proceeding further at this point \cite{AnomalyMatch}. 
\item The presence of unbroken flavor symmetries is guaranteed only in the ``magnetic" phase and the appearance of broken flavor symmetries in the ``electric" phase needs an additional check by the anomaly matching procedure.
\item There has been such a superpotential of this type constructed in SQCD with $N_f$=$N_c+1$, which satisfies all of consistency conditions imposed by the present theoretical knowledge \cite{AnomalyMatch}; therefore, this quantum superpotential, which is classically ill-defined, can correctly describe strongly coupled SQCD.
\end{enumerate}
It is, therefore, not so surprising that the instability of the SUSY gauge theories can be removed by generating Nambu-Goldstone superfields.  

Furthermore, if one admits the presence of the Higgs phase in any gauge theories with matters in fundamental representations of the gauge group, there is a corresponding confining phase smoothly connected from the Higgs phase \cite{SmoothComplementarity} by complementarity, the duality between the Higgs and confining phases \cite{tHooft,Susskind}.  Therefore, if one can consistently construct a superpotential describing physics in the confining phase and respecting the complementarity, a strongly coupled SUSY gauge theory can be characterized by ``electric" degrees of freedom instead of ``magnetic"  degrees of freedom.  Since it will be shown that the ``magnetic" description is not necessarily required to understand physics of the $SO(N_c)$ theory with $N_f=N_c-2$, which is to be described by an effective superpotential in terms of mesons (and baryons), it is of great importance to examine whether the ``magnetic" description is a must or not in other $SO(N_c)$ theories with $N_c-1\leq N_f$ $\leq$ $3(N_c-2)/2$
\footnote{The $SO(N_c)$ theories with $3(N_c-2)/2 < N_f < 3(N_c-2)$ are known to enter in an interacting Coulomb phase, where the well-understood $N$=2 duality can be transmitted to the $N$=1 duality.}.

In this paper, we would like to discuss dynamical properties of the SUSY $SO(N_c)$ gauge theories in order to demonstrate that strongly coupled SUSY gauge theories possess an ``electric" phase characterized by dynamical breakdown of flavor symmetries in addition to the conventional ``magnetic" phase based on the $N=1$ duality.  New results from the SUSY $SO(N_c)$ gauge theory with $N_f=N_c-1$ are included to show the appearance of flavor symmetry breaking and of holomorphic decoupling for $N_c-2 \leq N_f \leq 3(N_c-2)/2$.  The correct implementation of the instanton physics serves as one of our guiding principle to construct effective superpotentials, which is easily accomplished by employing $S$ \cite{VY}.  It is found that its consistent decoupling is realized by the existence of a flavor-singlet superfield orthogonal to the Nambu-Goldstone mode.  

We first show the evidence of the emergence of the flavor symmetry breaking in the ``electric" phase by constructing and examining our effective superpotential in the $SO(N_c) $ theory with $N_f=N_c-2$.  The flavor symmetry breaking of $SU(N_f)\rightarrow SO(N_f)$ is indeed induced by the instanton effects.  This flavor symmetry breaking is not respected by the $N=1$ duality.  The usual $N=1$ duality explains the physics of this $SO(N_c)$ theory by introducing two monopoles and an additional flavor-singlet chiral ``gauge" superfield \cite{SO}, which saturate all the anomalies arising from the unbroken flavor symmetries. The anomaly-matching is associated with the presence of the residual chiral $U(1)_R$ symmetry (shown in Table \ref{Tab:QuntumNumber})
 as an anomaly-free chiral $U(1)$ symmetry.  In our case, it is taken care of by the Nambu-Goldstone superfields and can be easily seen by invoking the old duality known as complementarity, which calls for the additional flavor singlet chiral ``gauge" superfield as well, corresponding to the $SO(2)$ gauge superfield contained in $SO(N_c)$ in the Higgs phase. 

The flavor symmetry is given by
\begin{equation}\label{Eq:FlavorSym}
 G = SU(N_f)\times U(1)_R,
\end{equation}
under which chiral quark superfields of $Q^i_A$ and chiral gauge superfields of $W_{[AB]}$ transform according to Table \ref{Tab:QuntumNumber}
\footnote{Throughout this paper, the labels of $i,j,\cdots$ specify the flavor indices of $SU(N_f)$ and those of $A,B,\cdots$ specify the color indices of $SO(N_c)$.}.
 We find the following patterns of dynamical flavor symmetry breaking:
\begin{equation}\label{Eq:FlavorSym0_N-2}
 G \rightarrow  SO(N_f)\times U(1)_R,
\end{equation}
for $N_f=N_c-2$, 
\begin{equation}\label{Eq:FlavorSym0_N-1}
 G \rightarrow  SO(N_f),
\end{equation}
for $N_f=N_c-1$ and $N_f=N_c$, and 
\begin{equation}\label{Eq:FlavorSym0_N}
 G \rightarrow  SO(N_c)\times SU(N_f-N_c)\times U(1)^\prime_R,
\end{equation}
for $N_f \geq N_c+1$, where $U(1)^\prime_R$ is a remnant of $U(1)_R$.  We also discuss the holomorphic decoupling property exhibited by a series of our effective superpotentials for $N_f\geq N_c-2$, which are continuously linked to those for $N_f < N_c-2$ discussed in Ref.\cite{SO_IS}, where dynamical flavor symmetry breaking occurs. 

We emphasize that, for $N_f\geq N_c-1$, where baryons are included, to find the spontaneous symmetry-breaking solution, we rely upon the plausible theoretical expectation that the SUSY theory has a smooth limit to the SUSY-preserving phase from the SUSY-breaking phase.  This expectation is based on the fact that the real physics is all lies in a SUSY breaking phase if there is SUSY at all.  Namely, any mathematically rigorous proof that resides on the exact SUSY loses its power when SUSY is broken.  Since any SUSY vacua are equally acceptable as ground states, to select a correct one among these vacua is only possible by examine physics with broken SUSY, which may be compared with the observed physics that indeed does not respect SUSY.  However, since such a non-supersymmetric deformation in a SUSY theory generally leads to qualitatively different types of physics from the SUSY physics, one has to check that the obtained SUSY vacuum in the SUSY limit of the non-supersymmetric vacuum possesses a consistent property with that of the SUSY theory. We provide two consistent checks: one is to examine that the proposed effective superpotential yields the predicted SUSY vacuum and the other is to examine that the expected physical spectrum realized on the SUSY vacuum respects the anomaly-matching property.  Once we find the location of the correct SUSY vacuum, we simply adopt this vacuum configuration at the startup and we can construct an effective theory residing on the SUSY vacuum without recourse to the non-supersymmetric deformation.

This paper is organized as follows: In Sec. \ref{sec:2}, effective superpotentials are constructed by following the classic method developed in Ref.\cite{VY,MPRV,YYY}.  In Sec. \ref{sec:3}-Sec. \ref{sec:5}, we discuss the dynamics of SUSY $SO(N_c)$ gauge theories based on the effective superpotentials proposed in Sec. \ref{sec:2}, respectively, for $N_f=N_c-2$, $N_f=N_c-1$ and $N_f\geq N_c$. In Sec. 

\section{Effective Superpotentials}\label{sec:2}
Our effective superpotentials consist of composite superfields made of quarks and gluons, which are mesons made of quarks:
\begin{equation}\label{Eq:FieldContentT}
 T^{ij}  =  \sum_{A=1}^{N_c} Q_A^iQ_A^j,  
\end{equation}
and a composite superfield made of two chiral gauge superfields:
\begin{equation}\label{Eq:FieldContentS}
S = \frac{g^2}{32\pi^2}\sum_{A,B=1}^{N_c} W_{[AB]}W_{[AB]},
\end{equation}
with the $SO(N_c)$ gauge coupling, $g$, where the gauge coupling is explicitly included in $S$ for the sake of the later discussions.  The inclusion of $S$ was advocated by Veneziano and Yankielowicz \cite{VY} some times ago while the formulation without $S$ has been motivated by Afleck, Dine and Seiberg \cite{EarlySeiberg}.  The role of $S$ is to reproduce the correct amount of the breaking of an anomalous $U(1)$ symmetry induced by instantons, which may not be physically required since this symmetry is not a conserved symmetry so that effective interactions need not respect its presence.  However, we adopt the description in terms of $S$ to evaluate the instanton effects by the effective superpotential approach.  In addition to these composites, there are specific baryons depending upon the number of colors and flavors, which are given by
\begin{itemize}
\item for $N_f=N_c-2$, a chiral flavor-singlet ``gauge" superfield made of $N_c-2$ quarks and a chiral gauge superfield:
\begin{eqnarray}\label{Eq:FieldContentN-2}
 {\tilde B} &=& 
\sum_{A_1\dots A_{N_c}}
\frac{1}{(N_c-2)!2!}
\varepsilon^{A_1{\dots}A_{N_c}}
Q_{A_1}^{i_1}\dots Q_{A_{N_c-2}}^{i_{N_c-2}}W_{[A_{N_c-1}A_{N_c}]},
\end{eqnarray}
\item for $N_f=N_c-1$, a ``diquark"-like composite made of $N_c-1$ quarks transforming as ${\bf N_c}$ of $SO(N_c)$:
\begin{eqnarray}\label{Eq:FieldContentN-1}
 d^A &=& 
\sum_{AA_1\dots A_{N_c}}
\frac{1}{(N_c-1)!}
\varepsilon^{AA_1{\dots}A_{N_c-1}}
Q_{A_1}^{i_1}\dots Q_{A_{N_c-1}}^{i_{N_c-1}},
\end{eqnarray}
which is confined to yield a color-singlet meson of
\begin{eqnarray}\label{Eq:FieldContentU}
B &=& \sum_{A=1}^{N_c} d^Ad^A,
\end{eqnarray}
\item for $N_f\geq N_c$, a baryon made of $N_c$ quarks:
\begin{equation}\label{Eq:FieldContentN}
 B^{[i_1i_2{\dots}i_{N_c}]} = 
\sum_{A_1\dots A_{N_c}}
\frac{1}{N_c!}
\varepsilon^{A_1{\dots}A_{N_c}}Q_{A_1}^{i_1}\dots Q_{A_{N_c}}^{i_{N_c}}.
\end{equation}
\end{itemize}
All these composite superfields respect the complementarity in our approach.  Especially, the first baryon is usually so selected to satisfy anomaly-matching \cite{SO_IS}, but now is required by the emergence of a corresponding field in the Higgs phase.

Our superpotential to be constructed in this section contains $f(Z)$ as a function of a new field $Z$ composed of the baryonic degrees of freedom for $N_f\geq N_c-1$, which is defined by 
\begin{equation}\label{Eq:FieldContentZN-1}
Z = \frac{B}{\det(T)},
\end{equation}
for $N_f=N_c-1$ and
\begin{eqnarray}\label{Eq:FieldContentZN}
Z &=& \frac{\sum_{i_1\cdots i_{N_f},j_1\cdots j_{N_f}} \varepsilon_{i_1\cdots i_{N_f}}\varepsilon_{j_1\cdots j_{N_f}}B^{[i_1\cdots i_{N_c}]}T^{i_{N_c+1}j_{N_c+1}}\cdots T^{i_{N_f}j_{N_f}}B^{[j_1\cdots j_{N_c}]}}{N_c!N_c!(N_f-N_c)!(N_f-N_c)!\det \left(T\right)},
\nonumber \\
\end{eqnarray}
for $N_f\geq N_c$, to be abbreviated to 
\begin{equation}\label{Eq:FieldContentZN_BB}
Z = \frac{BT^{N_f-N_c}B}{\det(T)}.
\end{equation}
These $Z$'s are neutral under the entire chiral symmetries including $U(1)_{anom}$ and the $Z$-dependence of $f(Z)$ cannot be determined by the symmetry principle.  For $N_f=N_c-2$, the baryonic ${\tilde B}$ needs not enter in $W_{\rm eff}$ through $f(Z)$.  Its dynamics turns out to be given by a $U(1)$ gauge theory with ${\tilde B}$ as a gauge field since ${\tilde B}$ turns out to be a massless chiral gauge superfield.  In the classical limit, it is readily observed that the new field $Z$ satisfies
\begin{equation}\label{Eq:ClassicalZ=1}
Z = 1,
\end{equation}
for $N_f\leq N_c$.  For $N_f> N_c$, there are no classical limits of superpotentials constructed out of $\det(T)$ and $Z$.  Instead of $Z$=1, it can be stated that
\begin{equation}\label{Eq:ClassicalZ=1ForNf_Nc}
\left(
\det(T)(T^{-1})^{N_f-N_c}
\right)^{i_ij_1\cdots i_{N_c}j_{N_c}} = B^{[i_1\cdots i_{N_c}]}B^{[j_1\cdots j_{N_c}]}.
\end{equation}
At the same time, $\det \left(T\right)$=0 follows.

The classic construction \cite{VY,MPRV,YYY} of effective superpotentials (for massless quarks) requires that not only they are invariant under the transformations of all the symmetries in $G$ of Eq.(\ref{Eq:FlavorSym}) but also they are compatible with the response from an anomalous $U(1)_{anom}$ symmetry, namely, $\delta{\cal L}$ $\sim$ $F^{\mu\nu}{\tilde F}_{\mu\nu}$, where ${\cal L}$ represents the lagrangian of the $SO(N_c)$ theory and $F^{\mu\nu}$ (${\tilde F}_{\mu\nu}\sim \varepsilon_{\mu\nu\rho\sigma}F^{\rho\sigma}$) is a gauge field strength. Its SUSY version applied to an effective superpotential, $W_{\rm eff}$, becomes
\begin{equation}\label{WeffAnomS}
\delta W_{\rm eff} = -2iN_f\alpha S,
\end{equation}
where $\alpha$ is the phase associated with $U(1)_{anom}$.  The invariance from $U(1)_R$ is stated as $\delta_R W_{\rm eff}=0$. Our effective superpotentials expressed in terms of $S$, $T$ and $f(Z)$ for the different sets of $N_f$ and $N_c$, should satisfy
\begin{eqnarray}
2\left( S\frac{\partial W_{\rm eff}}{\partial S}-W_{\rm eff}\right)
+
\frac{N_f-N_c-2}{N_f}
\left[
2T\frac{\partial W_{\rm eff}}{\partial T}+rN_c\frac{\partial W_{\rm eff}}{\partial B}
\right]
=
0,
\label{Eq:WeffEq}
\end{eqnarray}
from $\delta_R W_{\rm eff}$ and
\begin{eqnarray}
2T\frac{\partial W_{\rm eff}}{\partial T}+rN_c\frac{\partial W_{\rm eff}}{\partial B}
=
2N_fS
,
\label{Eq:WeffAnomEq}
\end{eqnarray}
from $\delta W_{\rm eff}$, where $r$=(2, 1) for ($N_f$=$N_c-1$, $N_f\geq N_c$).  

Considering the invariance from $SU(N_f)$, one can find that Eqs.(\ref{Eq:WeffEq}) and (\ref{Eq:WeffAnomEq}) are satisfied by $S\{\ln [S^{N_c-N_f-2}\det \left(T\right)/\Lambda^{3N_c-N_f-6}]+N_f-N_c+2\}$.  For $N_f \geq N_c-1$, it is obvious that $Z$ of Eqs.(\ref{Eq:FieldContentZN-1}) and (\ref{Eq:FieldContentZN}) satisfies $2T({\partial Z}/{\partial T})$ + $rN_c({\partial Z}/{\partial B})$ = 0. Furthermore, $Sh(Z)(=X)$ satisfies $S\partial X/{\partial S}-X=0$, where $h(Z)$ is any function of $Z$. It means that $Sh(Z)$ is a solution to Eq.(\ref{Eq:WeffEq}) and simultaneously serves as a general solution to Eq.(\ref{Eq:WeffAnomEq}).  If quark masses are included, we reach the following effective superpotential with $h(Z)$ = $\ln f(Z)$:
\begin{equation}\label{Eq:WeffN}
W_{\rm eff}=S 
\left\{ 
\ln\left[
\frac{
	S^{N_c-N_f-2}\det \left(T\right) f(Z)
}
{
	\Lambda^{3N_c-N_f-6}
} 
\right] 
+N_f-N_c+2\right\} - \sum_{i=1}^{N_f}m_iT^{ii},
\end{equation}
where $\Lambda$ is the scale of the $SO(N_c)$ theory and $m_i$ is the mass of the $i$-th quark.  It should be noted that $\det(T)$ cannot be classically defined for $N_f>N_c$ since $\det(T)$ simply described by quarks identically vanishes.  However, at the quantum level, it is possible to have mesons themselves as dynamical freedoms, which yield non-trivial configuration of $\det(T)$.  For $N_f\leq N_c-2$, our approach ends up with the effective superpotential given by
\footnote{The SUSY mass term will be omitted unless necessary}
\begin{equation}\label{Eq:WeffN-2}
W_{\rm eff}=S 
\left\{ 
\ln\left[
\frac{
	S^{N_c-N_f-2}\det \left(T\right)
}
{
	\Lambda^{3N_c-N_f-6}
} 
\right] 
+N_f-N_c+2\right\}
 - \sum_{i=1}^{N_f}m_iT^{ii}.
\end{equation}
For $N_f\leq N_c-3$, the superpotential of Eq.(\ref{Eq:WeffN-2}) for massless quarks is equivalent to the one discussed in Ref.\cite{SO_IS} after $S$ is eliminated.  Namely, one reaches
\begin{equation}\label{Eq:W_ADS_0}
W_{\rm eff}^{\prime}=-\epsilon_{N_c-N_f-2}\left( N_c-N_f-2\right) \left[ 
\frac{
	\Lambda^{3N_c-N_f-6}
}
{
	\det \left(T\right)
} 
\right]^{1/(N_c-N_f-2)},
\end{equation}
where $\epsilon_n$ (with $n=N_f-N_c+2$) is the n-th root of unity, which indicates that there is a SUSY minimum at $T\to\infty$, yielding no stable SUSY vacuum.  For $N_f\geq N_c-1$, by envisioning the decoupling of $S$, our superpotentials become
\begin{equation}\label{Eq:W_ADS}
W_{\rm eff}^{\prime}=\epsilon_{N_f-N_c+2}\left( N_f-N_c+2\right) \left[ 
\frac{
	\det \left(T\right) f(Z)
}
{
	\Lambda^{3N_c-N_f-6}
} 
\right]^{1/(N_f-N_c+2)},
\end{equation}
which indicates that the SUSY minimum is located at $\det \left(T\right) f(Z)=0$.

Let us comment on the use of $f(Z)$ in effective superpotentials \cite{UseOfZ}. It was first advocated in Ref.\cite{MPRV} to discuss properties of the $N=1$ SQCD with $N_f=N_c$. The appearance of $f(Z)$ was also encountered in Ref.\cite{Review} to demonstrate the power of symmetries and holomorphy.  For the reader's convenience, we repeat the discussions here.  The model is the simplest Wess-Zumino model with the tree-level superpotential, $W^{WZ}_{\rm tree}$, given by
\begin{equation}\label{Eq:WessZumino}
W^{WZ}_{\rm tree}=m\phi^2+h\phi^3, 
\end{equation}
where $\phi$, $m$ and $h$, respectively, have quantum numbers of (1, 1), ($-2$, 0) and ($-3$, $-1$) as two $U(1)$ charges of ($U(1)$, $U(1)_R$).  Since $h\phi/m$ is neutral under $U(1)\times U(1)_R$, the $Z$ field can be given by
\begin{equation}\label{Eq:WessZuminoZ}
Z=h\phi/m.
\end{equation}
The symmetries and holomorphy of the effective superpotential, $W^{WZ}_{\rm eff}$, restrict it to be of the form
\begin{equation}\label{Eq:WessZuminoW}
W^{WZ}_{\rm eff}=m\phi^2f(Z).
\end{equation}
This $W^{WZ}_{\rm eff}$ corresponds to Eqs.(\ref{Eq:WeffN}) and (\ref{Eq:WeffN-2}).  To find the explicit form of $f(Z)$ needs dynamical information of the model.  It is the requirement that this $W^{WZ}_{\rm eff}$ should coincide with  $W^{WZ}_{\rm tree}$ in the limit of $h\rightarrow 0$ and $m\rightarrow 0$ with arbitrary $g/m$, which yields $f(Z)=1+Z$. The same is true to determine $f(Z)$ of Eqs.(\ref{Eq:WeffN}) and (\ref{Eq:WeffN-2}).  This example shows that we need some dynamical information of the $SO(N_c)$ gauge theories to find the explicit form of $f(Z)$

The dynamical properties of our superpotentials for $N_f\geq N_c-1$ can be summarized as follows:
\begin{enumerate}
\item The classical limit of Eqs.(\ref{Eq:WeffN}) and (\ref{Eq:WeffN-2}) can be obtained by the same method in Ref.\cite{VY}. One can find the behavior of $W_{\rm eff}$ by invoking the definition of
\begin{equation}\label{Eq:Lambda}
\Lambda \sim \mu\exp (-8\pi^2/(3N_c-N_f-6)g^2), 
\end{equation}
where $\mu$ is a certain reference mass scale. In the classical limit of $g\rightarrow 0$, the resulting $W_{\rm eff}$ turns out to be:
\begin{eqnarray}\label{Eq:ClassicalS}
W_{\rm eff}
&{\mathop \to \limits_{g \to 0}}&
-S\ln\left(	\Lambda^{3N_c-N_f-6} \right)
=
-S\frac{(-8\pi^2)(3N_c-N_f-6)}{(3N_c-N_f-6)g^2}
\nonumber \\
&=& \frac{g^2}{32\pi^2} WW\frac{8\pi^2}{g^2}
=\frac{1}{4}WW,
\end{eqnarray}
which is nothing but the tree superpotential for the gauge kinetic term.
\item The SUSY minimum can be determined by the solutions to $\partial W_{\rm eff}/\partial S$=0 and $\partial W_{\rm eff}/\partial T^{ij}$=0, which are given by $S$=0 and $\det(T)f(Z)=0$ for $N_f\geq N_c-2$, yielding the vanishing nonperturbative contributions of Eq.(\ref{Eq:W_ADS}).  The solution with $\det(T)=0$ points to the origin of the moduli space in the ``magnetic" description while the solution with 
\begin{equation}\label{Eq:fZ=0}
f(Z)=0
\end{equation}
may allow 
\begin{equation}\label{Eq:det(T)=0}
\det(T)\neq 0, 
\end{equation}
in the ``electric" description with the Nambu-Goldstone superfields associated with $\det(T) \neq 0$
\footnote{
More severe constraints involving $f^\prime(Z)=0$ will be imposed on $f(Z)$ (See Eqs.(\ref{Eq:detT=0_N-1}) and (\ref{Eq:detT=0_Nc})).
}.
  Since $\det(T)\neq 0$, this vacuum is located away from the origin.  This is the case with $N_f=N_c-1$ and $N_f=N_c$.  Other cases with $N_f\geq N_c+1$ are based on $\det(T)= 0$ and $f(Z)=0$.  However, we find that the presence of ${\rm det}_{N_c}(T)\neq 0$, where ${\rm det}_{N_c}(T)$ is the determinant ranging over the $N_c\times N_c$ subspace of $N_f\times N_f$, indicates again the usefulness of the ``electric" description with the Nambu-Goldstone superfields.  It turns out that $f(Z)=0$ holds for $N_f\geq N_c-1$.
\end{enumerate}
From these observations, we find that the constraint of $f(Z)=0$ in the classical limit (stable for $N_f\leq N_c$) can take the simplest form given by
\begin{equation}\label{Eq:Classical_fZ}
f(Z) = (1-Z)^\rho ~ (\rho > 0),
\end{equation}
corresponding to the choice of $f(Z)=1+Z$ in the Wess-Zumino model.  This simplest form can also be applied to the cases with $N_f>N_c$ by the use of the holomorphic decoupling property.  This summary helps the readers understand why the flavor symmetry breaking is allowed to occur in our approach.  This is because $f(Z)$ is introduced in our analysis and the existence of the baryon degrees of freedom is essential.

In the $SO(N_c)$ theories with $N_f=N_c-4$ and $N_f=N_c-3$, it has been demonstrated in Ref.\cite{SO_IS} that there are two phase branches: one with the dynamically broken flavor symmetry of $SU(N_f)\rightarrow SO(N_c)$ and the other with the unbroken flavor symmetry.  In the symmetry broken phase, the $SO(N_c)$ theories have no vacuum.  This phase also exists for $N_f< N_c-4$ while the symmetry preserving phase exits for $N_f > N_c-3$.  In our superpotential approach relying upon Eqs.(\ref{Eq:WeffN}) and (\ref{Eq:WeffN-2}), the flavor symmetry breaking phase can also exist for $N_f > N_c-3$, where the cases with $N_f \geq N_c-1$ are based on the solution of $f(Z) = 0$ while the case with $N_f = N_c-2$ is based on the instanton physics, which is discussed in the next section.  

\section{${\bf N_f=N_c-2}$}\label{sec:3}
In this section, we show the direct evidence of the presence of dynamical flavor symmetry breaking, which is induced by the instanton of the $SO(N_c)$ theory with $N_f=N_c-2$.  The instanton is known to yield contributions from the gluinos and $N_f$ massless quarks, which can be specified by
\begin{equation}
\label{Eq:Instanton}
(\lambda\lambda)^{N_c-2}\det (\psi^i\psi^j), 
\end{equation}
where $\psi$ ($\lambda$) is a spinor component of $Q$ ($S$).  From Eq.(\ref{Eq:InstantonVEVMassiveNf=Nc-2}) in the Appendix \ref{Appendix:InstantonContributions}, we find that this instanton amplitude is equivalent to
\begin{equation}
\label{Eq:InstantonScalar}
\det (\phi^i\phi^j), 
\end{equation}
thus, yielding a SUSY preserving VEV as
\begin{equation}
\label{Eq:InstantonScalar2}
\det (\phi^i\phi^j) \sim \Lambda^{2N_f},
\end{equation}
where $\phi^i$ is the $i$-th scalar quark.  This evaluation clearly shows the appearance of the dynamical flavor symmetry breaking of
\begin{equation}
\label{Eq:FlavorSymBreaking}
SU(N_f) \times U(1)_R \rightarrow SO(N_f) \times U(1)_R,
\end{equation}
with $SO(N_f)$ as a maximal subgroup. 

On the other hand, our superpotential of Eq.(\ref{Eq:WeffN-2}) with $N_f=N_c-2$ becomes 
\begin{equation}\label{Eq:WeffX}
W_{\rm eff}=S 
\ln\left(
\frac{
	\det \left(T\right)
}
{
	\Lambda^{2N_f}
}
\right).
\end{equation}
The SUSY minimum is determined by requiring $\partial W_{\rm eff}/\partial \pi_{i,S}(\equiv W_{{\rm eff};i,S})$ to vanish:
\begin{equation}
W_{{\rm eff};i} = \frac{\pi_S}{\pi_i} = 0, \qquad
W_{{\rm eff};S}  = \ln\left(\prod\limits_{i = 1}^{N_f}\frac{\pi_i}{\Lambda^2}\right) = 0,
\label{Eq:Weff_iS}
\end{equation}
where  $\pi_i$ represents the scalar component of $T^{ii}$ and $\pi_S$ is the scalar component of $S$.  Other fields such as scalar components of $T^{ij}$ with $i\neq j$ can be set to vanish at the minimum; therefore, we omit these terms.  The solution to these equations is given by $\Pi_{i = 1}^{N_f}\pi_i=\Lambda^{2N_f}$ and $\pi_S$ = 0, leading to
\begin{equation}
\det \left(T\right) =\Lambda^{2N_f}, \quad S = 0.
\label{Eq:Solutions_iS}
\end{equation}
This vacuum is the same as Eq.(\ref{Eq:InstantonScalar}) generated by the instantons.  As a result, the maximal flavor symmetry respected by the $SO(N_c)$ theory turns out to be $SO(N_f) \times U(1)_R$ as in Eq.(\ref{Eq:FlavorSymBreaking}). In this end, the chiral Nambu-Goldstone superfields are produced as the low-energy degrees of freedom. This situation is similar to the one encountered in SQCD with $N_f=N_c$ \cite{Review,MPRV}. The corresponding superpotential in terms of $S$ is given by 
\begin{equation}\label{Eq:Weff_SQCD}
W_{\rm eff}=S 
\ln\left(
\frac{
	\det \left(T\right)-B_sB_s
}
{
	\Lambda^{2N_f}
}
\right),
\end{equation}
where $B_s$ and ${\bar B_s}$ are flavor-singlet chiral superfields (baryons) analogous to Eq.(\ref{Eq:FieldContentN}) without the flavor indices, respectively, made of quarks and anti-quarks.  The solution of
\begin{equation}
\det \left(T\right)-B_s{\bar B_s} = \Lambda^{2N_f}, \quad S = 0,
\label{Eq:Solutions_iS2}
\end{equation}
indicates the spontaneous flavor symmetry breaking.

The anomaly-matching property is seen by the use of the complementarity \cite{tHooft,Susskind}. The flavor symmetry breaking of $SU(N_f) \times U(1)_R$ is also realized in the Higgs phase by requiring $\langle 0 \vert\phi^i_A\vert 0 \rangle$ = $\delta^i_A\Lambda$ for $i,A$ = 1 $\sim$ $N_c-2$.  In the Higgs phase, the $SO(N_c)$ gauge symmetry is broken to the $SO(2)$ gauge symmetry. The low-energy degrees of freedom listed in Table \ref{Tab:COMP_N-2}
 are supplied by $Q^{\{ij\}}$ as the symmetric representation of $SO(N_f)$ and $W_{[AB]}$ as the chiral gauge superfields of the $SO(2)$ gauge symmetry.  These superfields directly correspond to the Nambu-Goldstone superfields of $T^{ij}$ (with Tr($T^{ij}$) = 0) associated with $SU(N_f)\rightarrow SO(N_f)$ and to the chiral flavor-singlet ``gauge" superfield of ${\tilde B}$. The decoupling of $S$ is achieved by the presence of a mass term formed by the flavor singlet field $T$ with Tr($T$)$\neq$0, which is not the Nambu-Goldstone mode.  The presence of $U(1)_R$ requires that the ``gauge" superfield of ${\tilde B}$ to be massless.  This masslessness of the chiral ``gauge" superfield of ${\tilde B}$ is the indication of the appearance of a gauged $U(1)$ symmetry \cite{SO_IS}.  Since the Nambu-Goldstone superfields are neutral under this $U(1)$ symmetry, this ``photon" does not interact with matter fields.  Our low-energy symmetry is found to be:
\begin{equation}\label{Eq:FlavorSymGauged}
 U(1)^{loc}\times SO(N_f)\times U(1)_R,
\end{equation}
where $U(1)^{loc}$ is associated with the massless composite gauge superfield.  This description of the $SO(N_c)$ gauge theory should be compared with the one in terms of ``magnetic" quarks \cite{SO_IS}.  The flavor symmetry breaking is absent and the low-energy degrees of freedom are supplied by two monopoles and the chiral ``gauge" superfield of ${\tilde B}$, whose anomalies match the original anomalies.  

Before closing this section, let us comment on the $SO(3)$ gauge theory.  Since our confining phase respects the constraint of $N_f\leq 3(N_c-2)/2$, $N_f=1$ is selected and the flavor symmetry is just $U(1)_R$.  In this case, our superpotential takes the simple form of 
\begin{equation}\label{Eq:WeffSO3}
W_{\rm eff}=S 
\ln\left(
\frac{
	T
}
{
	\Lambda^2
}
\right),
\end{equation}
where $T$ = $Q^1Q^1$, yielding $\langle T\rangle \neq 0$ at the SUSY minimum.  Since no flavor symmetry is broken, no Nambu-Goldstone superfields are generated.  In fact, the resulting would-be Nambu-Goldstone superfield, $T-\langle T\rangle$, forms a mass term with $S$ since $\partial^2W_{\rm eff}/\partial S\partial T \neq 0$ at $\langle T\rangle \neq 0$.  The physical particle is the massless composite gauge field of ${\tilde B}$ for $U(1)^{loc}$.  In the Higgs phase, by the VEV of $\langle\phi^1_A\rangle \propto \delta_{1A}$, $SO(3)\times U(1)_R$ breaks down to $SO(2)\times U(1)_R$ and the $SO(2)$ gauge field is the remaining massless particle.  The complementarity is satisfied as expected.  This Higgs picture has been adopted in the conventional description in terms of monopoles \cite{SO_IS}.  Therefore, both approaches start with the presence of the Higgs phase, where $SO(3)$ breaks down to $SO(2)(=U(1)^{loc})$.  The conventional approach introduces ``magnetic" degrees of freedom to explain the source of the singularity at $T$=0 while our approach uses the complementarity to find types of scalar condensation in the confining phase (for dynamical flavor symmetry breaking and the Nambu-Goldstone superfields as ``electric" degrees of freedom in general).

We are convinced that the $SO(N_c)$ gauge theory with $N_f=N_c-2$ allows the emergence of the dynamical flavor symmetry breaking in the ``electric" phase (for $N_c\neq$ 3).  We next would like to consider whether this phenomenon can also be realized in other $SO(N_c)$ gauge theories with $N_f\geq N_c-1$.

\section{${\bf N_f=N_c-1}$}\label{sec:4}
The effective superpotential takes the form of Eq.(\ref{Eq:WeffN}), which becomes
\begin{equation}\label{Eq:WeffN-1_0}
W_{\rm eff}=S 
\left\{ 
\ln\left[
\frac{
	\det \left(T\right) f(Z)
}
{
	S\Lambda^{2N_f-3}
} 
\right] 
+1\right\},
\end{equation}
and, equivalently,
\begin{equation}\label{Eq:WeffADSN-1}
W_{\rm eff}^\prime=
\frac{
	\det \left(T\right) f(Z)
}
{
	\Lambda^{2N_f-3}
},
\end{equation}
where $Z$ is given by
\begin{equation}\label{Eq:B_N-1}
Z=\frac{B}{\det(T)},
\end{equation}
as in Eq.(\ref{Eq:FieldContentZN}) and $B$ is the composite of $bb$ with the ``diquark", $b$, made of $N_c-1$ quarks as in Eq.(\ref{Eq:FieldContentU}).  Our SUSY minimum can be achieved at $\det(T)\neq 0$ and $f(Z)=0$.  However, it is not obvious that Eq.(\ref{Eq:WeffN-1_0}) leads to $\det(T)\neq 0$ because $\det(T)=0$ is also allowed at the SUSY minimum.

To understand that Eq.(\ref{Eq:WeffN-1_0}) really yields $\det(T)\neq 0$, we rely upon the plausible dynamical assumption that SUSY theories with the slightly broken SUSY are continuously connected with the SUSY theories in the SUSY limit.  As shown in the Appendix \ref{Appendix:BrokenSUSY}, we find that the flavor-singlet SUSY breaking does induce spontaneous flavor symmetry breakings of all the symmetries.  Therefore, the slightly broken $SO(N_c)$ theory with $N_f=N_c-1$ always provides $\det(T)\neq 0$.  In the limit of the unbroken SUSY, one can distinguish the vacuum with $\det(T)\neq 0$ from that with $\det(T)= 0$ by examining whether $\det(T)\neq 0$ obtained in the broken SUSY theories approaches to $\det(T)\neq 0$ or to $\det(T)= 0$ in the unbroken SUSY theories.

To deal with the SUSY breaking phase, our discussions are found to be more transparent by using the effective superpotential of Eq.(\ref{Eq:WeffN-1_0}) expressed in terms of $S$ than by using Eq.(\ref{Eq:WeffADSN-1}).  Since we are only concerned with the soft SUSY breaking but not with the soft flavor symmetry breaking, we adopt the simplest term that is invariant under the whole flavor symmetries, which can be easily accommodated by the following lagrangian, ${\cal L}_{mass}$, for the scalar quarks, $\phi_A^i$:
\begin{equation}\label{Eq:SUSYBreakingMass}
-{\cal L}_{mass}=\mu^2\sum_{i,A}\vert \phi_A^i \vert^2.
\end{equation}
Together with the potential terms arising from $W_{\rm eff}$, we find that
\begin{eqnarray}
& & V_{\rm eff}
=
G_T \bigg( 
   \sum_{i=1}^{N_f}  \left| W_{{\rm eff};i}\right|^2         
\bigg)
+ 
G_B\left| W_{{\rm eff};B}\right|^2
+
G_S \left| W_{{\rm eff};S}\right|^2
+
V_{\rm soft}, 
\label{Eq:VeffN-1} \\
& & V_{\rm soft}
=\mu^2\sum_{i=1}^{N_f} \left|\frac{\pi_i}{\Lambda_T}\right|^2
+
\mu^2
\left[
\left| \frac{\pi_B}{\Lambda^{2N_f-1}_B}\right|^2
+
\eta\left(
\frac{\pi_B}{\Lambda^{2(N_f-1)}_B}+\frac{\pi^\ast_B}{\Lambda^{2(N_f-1)}_B}
\right)
\right],
\label{Eq:VsoftN-1}
\end{eqnarray}
where $\Lambda_{T,B} \sim \Lambda$, $\pi_B$ denotes the scalar component of $B$, $\eta$ is a coefficient that measures the difference of the effective operators of $\pi_B$ and $\left| \pi_B\right|^2$ and the coefficient $G_T$ comes from the K$\ddot{\rm a}$hlar potential, $K$, which is assumed to be diagonal, $\partial^2 K/\partial T^{ik\ast}\partial T^{j\ell}$ = $\delta_{ij}\delta_{k\ell}G_T^{-1}$ with $G_T$ = $G_T(T^{\dagger}T)$, and similarly for $G_B$ = $G_B(B^{\dagger}B)$ and $G_S$ = $G_S(S^{\dagger}S)$.  Since we are interested in the SUSY-breaking phase in the vicinity of the SUSY-preserving phase, the leading terms of $\mu^2$ are sufficient to control the SUSY breaking effects.  Namely, we assume that $\mu^2/\Lambda^2\ll 1$.

Our SUSY $SO(N_c)$ dynamics at least allows one of the $\pi_i$ ($i$=1 $\sim$ $N_f$) to develop a VEV and let this be labeled by $i$ = $1$: $\left|\pi_1\right|$ = $\Lambda_T^2$. At this moment, $\det(T)=0$ is still satisfied.  The corresponding VEV at the SUSY broken vacuum is determined by solving  $\partial V_{\rm eff}/\partial\pi_i$ = 0, which is computed, in the Appendix \ref{Appendix:BrokenSUSY}, to be:
\begin{equation}\label{Eq:TijN-1}
G_TW_{{\rm eff};i}^\ast
\left( 1-\alpha \right) \frac{\pi_S}{\pi_i} 
 = \left( 1-\alpha \right) G_SW_{{\rm eff};S}^\ast +\beta X +M^2_T\left| \frac{\pi_i}{\Lambda_T}\right|^2, 
\end{equation}
for $i$=1${\sim}N_f$, where $\alpha$ and $\beta$ are defined in Eq.(\ref{Eq:alphabeta}). The SUSY breaking effect is specified by $\mu^2\vert \pi_1 \vert^2$ in Eq.(\ref{Eq:TijN-1}) through $M^2_T$ defined in Eq.(\ref{Eq:M2}) because of $\pi_1$ $\neq$ 0. From Eq.(\ref{Eq:TijN-1}) with Eq.(\ref{Eq:WeffPartial}), we find that  
\begin{equation}
\left| \frac{\pi_i}{\pi_1} \right|^2 = 1 + 
\frac{(M^2_T/\Lambda^2_T)(\left| \pi_1\right|^2-\left| \pi_i\right|^2)} {\left( 1-\alpha \right)G_SW_{{\rm eff};S}^\ast +\beta X +
(M^2_T/\Lambda^2_T)\left| \pi_i\right|^2}.
\label{Eq:VEVS_N-1}
\end{equation}
It is obvious that $\pi_{i\neq 1}$ = 0 cannot satisfy Eq.(\ref{Eq:VEVS_N-1}).  In fact, $\pi_{i\neq 1}$ = $\pi_1$ is a solution to this problem, leading to 
\begin{equation}
\left|\pi_{i=2\sim N_f}\right| = \left|\pi_1\right|\sim\Lambda^2. 
\label{Eq:VEVS_pi_1_N-1}
\end{equation}
As a result, all of the $\left|\pi_i\right|$ for $i$=$1\sim N_f$ dynamically acquire the same VEV once one of these $\left|\pi_i\right|$ receives a VEV.  Furthermore, by comparing Eq.(\ref{Eq:VeffPi_a_xi}) for $\pi_{i}$ with Eq.(\ref{Eq:VeffPi_B_xi}) for $\pi_B$, we observe that $\vert\pi_B\vert \sim \Lambda^{2N_f}$.

If this vacuum structure is valid even in the SUSY limit, we further find that, from Eq.(\ref{Eq:WeffPartial}) with $N=N_f=N_c-1$, the condition of $W_{{\rm eff};S}=0$ provides
\begin{equation}
f(z)=
\left(
\sum^{N_f-1}_{i=1}\frac{\Lambda^2}{\pi_i}
\right)
\frac{\pi_S}{\Lambda\pi_{N_f}}
\label{Eq:f(z)_N-1},
\end{equation}
at the SUSY minimum. By replacing $(1-\alpha)\pi_S/\pi_{N_f}$ with $W_{{\rm eff};N_f}$, we reach
\begin{equation}
\left(1-\alpha \right)f(z)=
\left(
\sum^{N_f-1}_{i=1}\frac{\Lambda^2}{\pi_i}
\right)
\frac{W_{{\rm eff};N_f}}{\Lambda},
\label{Eq:alpha_f(z)_N-1}
\end{equation}
at $z=1$.  Since $\pi_{i=1\sim N_f-1} \sim \Lambda^2$ as suggested by Eq.(\ref{Eq:VEVS_pi_1_N-1}) and $W_{{\rm eff};N_f}=0$ required for the SUSY minimum, we finally obtain the condition of the smooth SUSY limit on $f(z)$:
\begin{equation}
(1-\alpha )f(z)=f(z)-zf^\prime(z)=0,
\label{Eq:alpha_f(z)_N-1_no2}
\end{equation}
where Eq.(\ref{Eq:alphabeta}) for $\alpha$ is used.  Since $f(z)=0$ at the SUSY minimum, it requires that
\begin{equation}
f^\prime(z)=0.
\label{Eq:f(z)^prime=0_N-1}
\end{equation}
If the simplest form of $f(z)=(1-z)^\rho$ is taken, this constraint turns out to be
\begin{equation}
(1-z)^{\rho-1} \rightarrow 0,
\label{Eq:rho_f(z)_xi_N-1}
\end{equation}
as $z\rightarrow 1$, which requires that
\begin{equation}
\rho > 1.
\label{Eq:rhoN-1}
\end{equation}
Under this condition, we expect that the relation of $\left|\pi_{i=1\sim N_f}\right| \sim \Lambda^2$ is realized and yields $\det(T)\neq 0$. The $SO(N_c)$ theory with $N_f=N_c-1$ turns out to have the SUSY minimum with 
\begin{equation}
\det(T)\neq 0, \quad f(Z) = 0,\quad f^\prime(Z)=0.
\label{Eq:detT=0_N-1}
\end{equation}
The baryon of $B$ satisfies
\begin{equation}
B = \det(T),
\label{Eq:detT=B_N}
\end{equation}
leading to $B \neq 0$. The $SO(N_c)$ theory generates the dynamical flavor symmetry breaking of
\begin{equation}
SU(N_f)\times U(1)_R \rightarrow SO(N_f),
\label{Eq:SYM_N-1}
\end{equation}
as expected. 

Since the residual flavor symmetry of $SO(N_c)$ is anomaly-free, we do not worry about the anomaly-marching.  Our low-energy spectrum contains the Nambu - Goldstone superfields that have the direct correspondence with the massless particles in the Higgs phase. In the Higgs phase, this dynamical breakdown of $SU(N_f) \times U(1)_R$ can be realized by requiring that $\langle 0| \phi_A^a|0\rangle$ = $\delta_A^a\Lambda_T$ for $a,A$ = 1 $\sim$ $N_c-1$.  As shown in Table \ref{Tab:COMP_N-1}, the both spectra are precisely coincident with each other.  The decoupling of $S$ can be achieved by the presence of the field with the flavor-singlet linear combination of Tr$(T)$ and $B$ orthogonal to the Nambu-Goldstone mode that forms a mass term with $S$.  Conversely, the Nambu-Goldstone mode is the leftover piece that fails to form a mass term with $S$.  Therefore, the existence of $B$ is essential to provide the consistent decoupling of $S$.

The non-supersymmetric deformation indicates that $\det(T)\neq 0$ and $B\neq 0$. It is not obvious that this vacuum configuration reflects the correct SUSY vacuum.  We have shown that this non-supersymmetric vacuum becomes a correct SUSY vacuum that can be produced by our ``supersymmetric" superpotential only if $f^\prime(Z)=0$ is satisfied and that this SUSY vacuum respects the anomaly-matching property dynamically supported by the resulting Nambu-Goldstone superfields.

\section{${\bf N_f\geq N_c}$}\label{sec:5}
The effective superpotential in this case becomes
\begin{equation}\label{Eq:WeffN_0}
W_{\rm eff}=S 
\left\{ 
\ln\left[
\frac{
	S^{N_c-N_f-2}\det \left(T\right) f(Z)
}
{
	\Lambda^{3N_c-N_f-6}
} 
\right] 
+N_f-N_c+2\right\},
\end{equation}
and, equivalently,
\begin{equation}\label{Eq:WeffADSN}
W_{\rm eff}^\prime=\left( N_f-N_c+2\right) \left[
\frac{
	\det \left(T\right) f(Z)
}
{
	\Lambda^{3N_c-N_f-6}
}
\right]^{1/\left( N_f-N_c+2\right)},
\end{equation}
where $Z$ is given by
\begin{equation}\label{Eq:B_N}
Z=\frac{BT^{N_f-N_c}B}{\det(T)},
\end{equation}
as in Eq.(\ref{Eq:FieldContentZN_BB}) and $B$ is the composite made of $N_c$-quarks as in Eq.(\ref{Eq:FieldContentN}).  The SUSY minimum is realized at $\det \left(T\right) f(Z)$ = 0.  For $N_f>N_c$, the requirement of $\det \left(T\right)$ = 0 is consistent with the classical limit.  It will be shown that ${\rm det}_{N_c}(T)\neq 0$ is obtained and that it certainly satisfies $\det \left(T\right)$ = 0.  The condition of ${\rm det}_{N_c}(T)\neq 0$ also ensures the presence of the well-defined $Z$ because at the SUSY minimum, 
\begin{equation}\label{Eq:Valide_Z}
\langle Z\rangle = \frac{\langle BT^{N_f-N_c}B\rangle}{\langle \det(T)\rangle} = \frac{\langle B^{[1,...,N_c]}B^{[1,...,N_c]}\rangle}{\prod_{i=1}^{N_c}\pi_i}
\end{equation}
becomes well-defined if $\langle B^{[1,...,N_c]}\rangle\neq 0$, where $\langle T^{ij}\rangle=\delta_{ij}\pi_i$ ($i=1\sim N_c$) and others = 0.

With the flavor-singlet SUSY breaking accommodated by Eq.(\ref{Eq:SUSYBreakingMass}), the potential is given by
\begin{eqnarray}
& & V_{\rm eff}
=
G_T \bigg( 
   \sum_{i=1}^{N_f}  \left|W_{{\rm eff};i}\right|^2         
\bigg)
+ 
G_B\left| W_{{\rm eff};B}\right|^2
+
G_S \left| W_{{\rm eff};S} \right|^2
+
V_{\rm soft}, 
\label{Eq:VeffN} \\
& & V_{\rm soft}
=\mu^2\sum_{i=1}^{N_f} \left|\frac{\pi_i}{\Lambda_T}\right|^2
+
\mu^2
\left| \frac{\pi_B}{\Lambda^{2N_f-1}_B}\right|^2,
\label{Eq:VsoftN}
\end{eqnarray}
where $\mu^2/\Lambda^2\ll 1$.  We repeat the same discussions as those in Sec. 
\begin{equation}\label{Eq:VEVS_N}
\left| \frac{\pi_i}{\pi_1} \right|^2 = 1 + 
\frac{(M^2_T/\Lambda^2_T)(\left| \pi_1\right|^2-\left| \pi_i\right|^2)} {\left( 1-\alpha \right)G_SW_{{\rm eff};S}^\ast +\beta X +
(M^2_T/\Lambda^2_T)\left| \pi_i\right|^2},
\end{equation}
but for $i=1\sim N_c$ instead of $i=1 \sim N_f$.  Again, it is obvious that $\pi_{i\neq 1}$ = 0 cannot satisfy Eq.(\ref{Eq:VEVS_N}) and the solution is $\left|\pi_{i=2\sim N_c}\right|$ = $\left|\pi_1\right|$ ($\sim\Lambda^2$), which also indicates $\pi_B\sim \Lambda^{N_c}$.  From Eq.(\ref{Eq:WeffPartial}) with $N=N_c\leq N_f$, we find that the condition of the smooth SUSY limit on $f(z)$ is provided by
\begin{equation}
f(z)=
\sum^{N_c-2}_{i=1}\frac{\Lambda^2}{\pi_i}
\cdot
\sum^{N_f}_{i=N_c-1}\frac{\pi_S}{\Lambda\pi_i},
\label{Eq:f(z)_N}
\end{equation}
which is transformed into
\begin{equation}
\left(1-\alpha \right)^2f(z)=
\sum^{N_c-2}_{i=1}\frac{\Lambda^2}{\pi_i}
\cdot
\sum^{N_f}_{i=N_c-1}\frac{W_{{\rm eff};i}}{\Lambda}.
\label{Eq:f(z)_N_2}
\end{equation}

The smooth SUSY limit, giving $W_{{\rm eff};i} = 0$, for the suggested minimum of $\pi_{i=1\sim N_c} \sim \Lambda^2$, is realized if
\begin{equation}
\left(1-\alpha \right)^2f(z)=\frac{(f(z)-zf^\prime (z))^2}{f(z)}=0,
\label{Eq:alpha_f(z)_N_no2}
\end{equation}
which becomes
\begin{equation}
\frac{f^\prime (z)^2}{f(z)}=0,
\label{Eq:f(z)^prime=0_N}
\end{equation}
for $f(z)=0$.  By using the simplest form of $f(z)=(1-z)^\rho$, we find that
\begin{equation}
(1-z)^{\rho-2} \rightarrow 0,
\label{Eq:rho_f(z)_xi_N}
\end{equation}
as $z\rightarrow 1$, which leads to
\begin{equation}
\rho > 2.
\label{Eq:rhoN}
\end{equation}

The SUSY-preserving phase can have the smooth limit from the SUSY-broken theory with its minimum given by $\left|\pi_{i=1\sim N_c}\right|\sim\Lambda^2$.  It turns out that the SUSY minimum is characterized by
\begin{equation}
{\rm det}_{N_c}(T)\neq 0, \quad f(Z)=0, \quad \frac{f^\prime (Z)^2}{f(Z)}=0.
\label{Eq:detT=0_Nc}
\end{equation}
The baryon of $B$ satisfies the quantum constraint of
\begin{equation}
\det(T) = BT^{N_f-N_c}B.
\label{Eq:detT=BTB_N}
\end{equation}
Then, we are left with the dynamical flavor symmetry breaking of
\begin{equation}
SU(N_f)\times U(1)_R \rightarrow SO(N_f),
\label{Eq:SYM_N}
\end{equation}
for $N_f = N_c$, and
\begin{equation}
SU(N_f)\times U(1)_R \rightarrow SO(N_f)\times SU(N_f-N_c)\times U(1)^\prime_R.
\label{Eq:SYM_N_2}
\end{equation}
for $N_f \geq N_c+1$, where $U(1)^\prime_R$ is defined in Table \ref{Tab:COMP_N}. 

This dynamical breaking can really persist in the SUSY theories only if the anomaly-matching property is respected for $N_f\geq N_c+1$. The dynamical breakdown of $SU(N_f) \times U(1)_R$ can also be realized by the corresponding Higgs phase defined by $\langle 0| \phi_A^a|0\rangle$ = $\delta_A^a\Lambda_T$ for $a,A$ = 1 $\sim$ $N_c$. The anomaly-matching is trivially satisfied in the Higgs phase, whose spectrum is found to be precisely equal to that of the massless composite Nambu-Goldstone superfields represented by $T^{ab}$ with Tr($T^{ab}$) = 0, $T^{ia}$ and a linear combination of Tr$_{N_c}(T$) and $B^{[12\cdots N_c]}$, where $a,b=1\sim N_c$, $i=N_c+1\sim N_f$ and Tr$_A(T)=\sum^{N_c}_{A=1}T^{AA}$.  These are listed in Table \ref{Tab:COMP_N}.  Our superpotential, thus, assures that the anomaly-matching is a dynamical consequence of the symmetry breakdown.  No other composite fields such as those including gauge chiral superfields are present in the low-energy spectrum.    For $N_f = N_c$, no anomaly-matching is required but the low-energy degrees of freedom are represented by the Nambu-Goldstone superfields.  The decoupling of $S$ is achieved by the presence of the flavor-singlet linear combination of Tr$_{N_c}(T)$ and $B^{[12\cdots N_c]}$, which is orthogonal to the Nambu-Goldstone mode, that forms a mass term with $S$.  Again,  as in the case of $N_f=N_c-1$, the existence of $B^{[12\cdots N_c]}$ is essential to provide the consistent decoupling of $S$.

\section{Holomorphic Decoupling}\label{sec:6}
In this section, we demonstrate that the holomorphic decoupling works in our proposed effective superpotentials.  Let us first discuss the holomorphic decoupling property of Eq.(\ref{Eq:WeffN}) for $N_f\geq N_c$.  The property can be seen by adding a mass of $m_{N_f}$ to the $N_f$-th quark, yielding
\begin{equation}\label{Eq:WeffMassN}
W_{\rm eff}=S 
\left[ 
\ln
\left(
\frac{
S^{N_c-N_f-2}\det \left(T\right)f(Z)
}
{
	\Lambda^{3N_c-N_f-6}
}
\right)
+N_f-N_c+2
\right]
-m_{N_f}T^{N_fN_f}.
\end{equation}
At the SUSY minimum, 
\begin{equation}\label{Eq:KonishiAnomaly}
S=m_{N_f}T^{N_fN_f}
\end{equation}
is derived from $W_{{\rm eff};i=N_f}=0$, which is known as the Konishi anomaly relation \cite{KonishiAnomaly}. During the course of the decoupling of $T^{N_fN_f}$, the field, $T$, can be divided into ${\tilde T}$ with a light flavor $(N_f-1)$ $\times$ $(N_f-1)$ submatrix and $T^{N_fN_f}$ and also $B$ into light flavored ${\tilde B}$ and heavy flavored parts. The off-diagonal elements of $T$ and the heavy flavored $B$ vanish at the SUSY minimum.  Inserting these relations into Eq.(\ref{Eq:WeffMassN}),  we finally obtain
\begin{equation}\label{Eq:WeffMassN2}
W_{\rm eff}=S 
\left[ 
\ln
\left(
\frac{
S^{N_c-N_f-1}\det ({\tilde T})f({\tilde Z})
}
{
	{\tilde \Lambda}^{3N_c-N_f-5}
}
\right)
+N_f-N_c+1
\right]
,
\end{equation}
which is nothing but Eq.(\ref{Eq:WeffMassN}) by letting $N_f\rightarrow N_f-1$ with $T^{N_fN_f}$ decoupled, where ${\tilde Z}$ = ${\tilde B}{\tilde T}^{N_f-N_c-1}{\tilde B}$/${\rm det}({\tilde T})$ from $Z$ = ${\tilde B}T^{N_fN_f}{\tilde T}^{N_f-N_c-1}{\tilde B}/T^{N_fN_f}{\rm det}({\tilde T})$ and ${\tilde \Lambda}^{3N_c-N_f-5}$ = $m\Lambda^{3N_c-N_f-6}$.  This decoupling is successively applied to the $SO(N_c)$ theory until $N_f$ is reduced to $N_c$. Therefore, we observe that the decoupling is achieved by our effective superpotentials with $N_f \geq N_c$.

Next, we start with the effective superpotential for $N_f=N_c$:
\begin{equation}\label{Eq:WeffMassNf=Nc}
W_{\rm eff}=S 
\left[ 
\ln
\left(
\frac{
\det \left(T\right) f(Z)
}
{
	S^2\Lambda^{2N_c-6}
}
\right)
+2
\right]
-m_{N_f}T^{N_fN_f}.
\end{equation}
After the decoupling of $T^{N_fN_f}$ is completed, the baryon of $B$ made of $N_f$ quarks ceases to exist.  This baryon described by
\begin{equation}\label{Eq:BforNf=Nc}
B^{[1\cdots N_f]} \sim \varepsilon^{A_1\cdots A_{N_c-1}A_{N_c}}Q^1_{A_1}\cdots Q^{N_f-1}_{A_{N_c-1}}Q^{N_f}_{A_{N_c}}
\end{equation}
can be formally transformed into a color ${\bf N_c}$ baryon just with the $N_f$-th quark decoupled:
\begin{equation}\label{Eq:BforNf=Nc-1}
B^{A_{N_c}} \sim \varepsilon^{A_1\cdots A_{N_c-1}A_{N_c}}Q^1_{A_1}\cdots Q^{N_f-1}_{A_{N_c-1}},
\end{equation}
which is nothing but our ``diquark"-like composite $b^A$ of Eq.(\ref{Eq:FieldContentN-1}).  Namely, the $Z$ field has a smooth transition of 
\begin{equation}\label{Eq:BforNf=NcToBforNf=N_c-1}
\frac{B^{[1\cdots N_f]}B^{[1\cdots N_f]}}{\det(T)} 
\rightarrow 
\frac{b^Ab^AT^{N_fN_f}}{\det({\tilde T})T^{N_fN_f}}
=
\frac{b^Ab^A}{\det({\tilde T})}
\left(
\equiv \frac{B}{\det({\tilde T})}
\right).
\end{equation}
In this case, we obtain
\begin{equation}\label{Eq:WeffDecopledNf=Nc-1}
W_{\rm eff}=S 
\left[ 
\ln
\left(
\frac{
\det ({\tilde T}) f({\tilde Z})
}
{
	S{\tilde \Lambda}^{2N_c-5}
}
\right)
+1
\right],
\end{equation}
where ${\tilde Z}=B/\det({\tilde T})$ and ${\tilde \Lambda}^{2N_c-5} = m_{N_fN_f}\Lambda^{2N_c-6}$.  This effective superpotential is nothing but the one for $N_f=N_c-1$ given by Eq.(\ref{Eq:WeffN-1_0})
\footnote{
Since $W_{\rm eff}$ obtained from the decoupling from the one with $N_f=N_c$ has $\rho > 2$ for $f(Z)=(1-Z)^\rho$, $W_{\rm eff}$ with $N_f=N_c-1$ is also characterized by $\rho > 2$ instead of $\rho > 1$ as in Eq.(\ref{Eq:rhoN-1}) if the decoupling property is implemented.}.
Therefore, our holomorphic decoupling property is well described by introducing this ``diquark"-like composite, which is also required to provide $f(Z)=0$ instead of $\det(T)=0$ in the $N=1$ duality.

Having the effective superpotential of Eq.(\ref{Eq:WeffN-1_0}) for $N_f=N_c-1$ in our hand by the decoupling procedure, we examine the decoupling of the $N_f$-th quark, which similarly suggests $B^{[AB]}B^{[AB]}$ ($=B^\prime$)), where $B^{[AB]}$ supplied from $B^A$ with the $N_f$-th quark decoupled.  In the target case of $N_f=N_c-2$, the low energy degrees of freedom can be supplied by the Nambu-Goldstone superfields alone that also produce the required anomalies together with the baryonic ${\tilde B}$.  Furthermore, the consistent decoupling of $S$ is achieved in the spectrum without $B^\prime$ unlike in other cases. Therefore, there is no dynamical reason that calls for $B^\prime$.  Namely, $B^\prime$ is heavy if it exists at all. Instead, $B^{[AB]}$ together with the gauge superfield, $W_{[AB]}$, is confined to form the chiral flavor-singlet ``gauge" superfield of ${\tilde B}$.  Without $B^\prime$, we reach
\begin{equation}\label{Eq:WeffDecopledNf=Nc-2}
W_{\rm eff}=S
\ln
\left(
\frac{
\det ({\tilde T})
}
{
	{\tilde \Lambda}^{2N_c-4}
}
\right),
\end{equation}
where ${\tilde \Lambda}^{2N_c-4} = m_{N_fN_f}\Lambda^{2N_c-5}$, which is the effective superpotential of Eq.(\ref{Eq:WeffX}) for $N_f=N_c-2$.

The same procedure to reduce $N_f$ from $N_f=N_c-2$ yields
\begin{equation}\label{Eq:WeffDecopledNf=Nc-3}
W_{\rm eff}=S
\left[
\ln
\left(
\frac{
S\det ({\tilde T})
}
{
	{\tilde \Lambda}^{2N_c-3}
}
\right)-1
\right],
\end{equation}
where ${\tilde \Lambda}^{2N_c-3} = m_{N_fN_f}\Lambda^{2N_c-4}$, which is the same as 
\begin{equation}\label{Eq:WeffDecopledNf=Nc-3_2}
W_{\rm eff}=S 
\left[ 
\ln
\left(
\frac{
S^{N_c-N_f-2}\det \left(T\right)
}
{
	\Lambda^{3N_c-N_f-6}
}
\right)
+N_f-N_c+2
\right],
\end{equation}
for $N_f=N_c-3$.  It is further converted into
\begin{equation}\label{Eq:WeffDecopledNf=Nc-3_3}
W_{\rm eff}^{\prime}=-\epsilon_{N_c-N_f-2}\left( N_c-N_f-2\right) \left[ 
\frac{
	\Lambda^{3N_c-N_f-6}
}
{
	\det \left(T\right)
} 
\right]^{1/(N_c-N_f-2)},
\end{equation}
which is the one realized in the branch with the spontaneous flavor symmetry breaking for $N_f \leq N_c-3$ \cite{SO_IS}.

Our proposed effective superpotentials turn out to possesses the holomorphic decoupling property provided that
\begin{enumerate}
\item for $N_f\geq N_c$, the baryonic composites of the form of $\varepsilon^{A_1\cdots A_{N_c-1}A_{N_c}}Q^1_{A_1} Q^2_{A_2}$ $\cdots$ $Q^{N_c-1}_{A_{N_c-1}}Q^{N_c}_{A_{N_c}}$ take care of the presence of $f(Z)$,
\item for $N_f = N_c-1$, the composite of the form of $B = b^Ab^A$ with the baryonic $b^A= \varepsilon^{AA_1\cdots A_{N_c-1}}Q^1_{A_1} Q^2_{A_2}\cdots Q^{N_f}_{A_{N_c-1}}$ takes care of the presence of $f(Z)$,
\item for $N_f\leq N_c-2$, no baryonic composites are activated and $f(Z)$ disappears.
\end{enumerate}

\section{Instanton Effects for ${\bf N_f\geq N_c-1}$}\label{sec:7}
Since our effective superpotentials respect the holomorphic decoupling property, the $SO(N_c)$ theories with $N_f \geq N_c-1$ are reduced, by successive decoupling, to the $SO(N_c)$ theory with $N_f=N_c-2$, where the instanton effects are properly taken into account.  So, the instanton effects seem to be obviously included in the effective superpotentials with $N_f \geq N_c-1$.  However, it should be noted that the instanton effects in the $SO(N_c)$ theory with more than $N_c-2$ massless quarks do not affect the location of the SUSY minimum.  The decoupling from the case with $N_f> N_c-2$ do not care about the instanton contributions.  Namely, the massless $SO(N_c)$ physics with $N_f \geq N_c-1$, where the instanton contributions are irrelevant, differs from the partially massive $SO(N_c)$ theory with $N_f-N_c+2$ massive quarks, where the instanton contributions are relevant.  Therefore, the massless limit of the partially massive $SO(N_c)$ theory does not reproduce the corresponding massless $SO(N_c)$ physics.  It is, thus, not obvious that our effective superpotentials correctly include the instanton contributions.

The massless $SO(N_c)$ theory yields the instanton contributions from the gauginos and $N_f$ quarks, which are expressed as
\begin{equation}
\label{Eq:InstantonN}
(\lambda\lambda)^{N_c-2}\det (\psi^i\psi^j).
\end{equation}
In the case with $N_f-N_c+2$ massive quarks labeled by $i$=$N_c-1$ $\sim$ $N_f$, from Eq.(\ref{Eq:InstantonVEVMassive}) in the Appendix \ref{Appendix:InstantonContributions}, we find that this instanton amplitude is further transformed into \cite{MassiveInstanton}:
\begin{equation}
\label{Eq:InstantonNScalar}
{\rm det}_{N_c-2}(\phi^i\phi^j),
\end{equation}
which is allowed to become constant while SUSY is kept unbroken.  Therefore, Eq.(\ref{Eq:InstantonNScalar}) leads to
\begin{eqnarray}
&&
\langle \prod^{N_c-2}_{i=1}\phi_i\phi_i\rangle
\sim
\Lambda^{2(N_c-2)}\prod^{N_f}_{i=N_c-1}\frac{m_i}{\Lambda},
\label{Eq:MassiveInstanton0}
\end{eqnarray}
which is equivalent to
\begin{equation}
\label{Eq:MassiveInstanton}
\prod_{i=1}^{N_c-2}\pi_i = c\Lambda^{2N_c-4}\prod_{i=Nc-1}^{N_f}\frac{m_i}{\Lambda}, 
\end{equation}
where $c$ is a non-vanishing coefficient.

On the other hand, our superpotentials provide the corresponding SUSY minimum derived by
\begin{equation}
\label{Eq:SUSYminimum}
\frac{\partial W_{\rm eff}}{\partial \pi_i} = 0, \quad
\frac{\partial W_{\rm eff}}{\partial \pi_S} = 0,
\end{equation}
which give our constraints:
\begin{equation}
\label{Eq:SUSYmin_a}
\left( 1-\alpha \right)\frac{\pi_S}{\pi_i} = m_i,
\end{equation}
for the massive flavors of $i=N_c-1\sim N_c$, 
\begin{equation}
\label{Eq:SUSYmin_i}
\frac{\pi_S}{\pi_i} = m_i,
\end{equation}
for the massive flavors of $i=N_c+1\sim N_f$, and
\begin{equation}
\label{Eq:SUSYmin_S}
\ln
\left(
\sum^{N_c-2}_{i=1}\frac{\pi_i}{\Lambda^2}
\cdot
\sum^{N_f}_{i=N_c-1}\frac{\Lambda\pi_i}{\pi_S}
\cdot
f(z)
\right)
=0.
\end{equation}
Inserting Eqs.(\ref{Eq:SUSYmin_a}) and (\ref{Eq:SUSYmin_i}) into Eq.(\ref{Eq:SUSYmin_S}) gives
\begin{equation}
\label{Eq:SUSYmin_constraintN}
\left( 1-\alpha\right)^2 f(z) = 
\sum^{N_c-2}_{i=1}\frac{\Lambda^2}{\pi_i}
\cdot
\sum^{N_f}_{i=N_c-1}\frac{m_i}{\Lambda},
\end{equation}
for $N_f \geq N_c$ and
\begin{equation}
\label{Eq:SUSYmin_constraintN-1}
\left( 1-\alpha\right) f(z) = 
\sum^{N_c-2}_{i=1}\frac{\Lambda^2}{\pi_i}
\cdot
\frac{m_{N_c-1}}{\Lambda},
\end{equation}
for $N_f = N_c-1$.

Our result of Eqs.(\ref{Eq:SUSYmin_constraintN}) and (\ref{Eq:SUSYmin_constraintN-1}) coincides with Eq.(\ref{Eq:MassiveInstanton}) from the instantons if 
\begin{equation}
\label{Eq:Instanton_vs_ours}
\left(1-\alpha\right)^nf(z) = \frac{1}{c}
\end{equation}
with $n=1$ ($n=2$) for $N_f=N_c-1$ ($N_f\geq N_c$), is satisfied.  Namely, the possible mass-dependence of $f(z)$ found in Eqs.(\ref{Eq:SUSYmin_constraintN}) and (\ref{Eq:SUSYmin_constraintN-1}) is completely cancelled.  The important constraint of
\begin{equation}
\label{Eq:f(z)!=0}
f(z)\neq 0
\end{equation}
is required in this partially massive $SO(N_c)$ theory.  For the simplest case of $f(z)=(1-z)^\rho$, it leads to
\begin{equation}
\label{Eq:z!=1}
z\neq 1,
\end{equation}
indicating that the classical constraint characterized by $z=1$ is modified in the $SO(N_c)$ dynamics with the $N_f-N_c+2$ massive quarks. Therefore, the massless limit, where all of $N_f-N_c+2$ massive quarks become massless, does not reproduce the corresponding massless theory, where $f(z)$=0 is preserved.

Since $f(z)\neq 0$, Eqs.(\ref{Eq:SUSYmin_constraintN}) and (\ref{Eq:SUSYmin_constraintN-1}) provides
\begin{equation}
\label{Eq:pi_i!=0}
\pi_{i=1\sim N_c-2} \neq 0.
\end{equation}
and the Konishi anomaly relation dictates 
\begin{equation}
\label{Eq:pi_S=0}
\pi_S=\pi_{i=Nc-1\sim N_f}=0.
\end{equation}
In other words, in the effective superpotential without $S$ given by
\begin{eqnarray}\label{Eq:W_ADS_massive}
W_{\rm eff}^{\prime}&=&\epsilon_{N_f-N_c+2}\left( N_f-N_c+2\right) \left[ 
\frac{
	\det \left(T\right) f(Z)
}
{
	\Lambda^{3N_c-N_f-6}
} 
\right]^{1/(N_f-N_c+2)}
\nonumber \\
\quad
&& - \sum_{i=N_c-1}^{N_f}m_iT^{ii}
,
\end{eqnarray}
the SUSY minimum is, in fact,  realized at
\begin{equation}
\label{Eq:SUSY_massive}
{\rm det}_{N_c}(T) \neq 0,\quad f(Z) \neq 0, \quad T^{ii}=0,
\end{equation}
for $i=N_c-1\sim N_f$, as determined from Eqs.(\ref{Eq:pi_i!=0}) and (\ref{Eq:pi_S=0}). Therefore, we conclude that our superpotentials describe the correct flavor symmetry breaking with the one induced by the instantons of the $SO(N_c)$ theories in the case that $N_f-N_c+2$ quarks get massive while $N_c-2$ quarks remain massless.  

\section{Summary}\label{sec:8}
We have shown that, in the $SO(N_c)$ theory with $N_f=N_c-2$, the instanton effects in the ``electric" phase generate the flavor symmetry breaking, which cannot be provided by the $N=1$ description in terms of two monopoles .  The proposed effective superpotential of
\begin{equation}\label{Eq:SummaryWeffX}
W_{\rm eff}=S 
\ln\left(
\frac{
	\det \left(T\right)
}
{
	\Lambda^{2N_f}
}
\right),
\end{equation}
requires
\begin{equation}
\det \left(T\right) =\Lambda^{2N_f},
\label{Eq:SummarySolutions_iS}
\end{equation}
which defines the SUSY minimum identical to the one determined by the instantons.  The similar minimum has been already encountered in SQCD with $N_f=N_c$, which is based on the use of the baryon of $B_s$ in
\begin{equation}\label{Eq:SummaryWeff_SQCD}
W_{\rm eff}=S 
\ln\left(
\frac{
	\det \left(T\right)-B_sB_s
}
{
	\Lambda^{2N_f}
}
\right),
\end{equation}
and the constraint of 
\begin{equation}
\det \left(T\right)-B_s{\bar B_s} = \Lambda^{2N_f}
\label{Eq:SummrySolutions_iS2}
\end{equation}
is obtained and also indicates the flavor symmetry breaking.  So, there is no positive reason that excludes the possibility to use $\det (T)\neq 0$ in the $SO(N_c)$ theory with $N_f=N_c-2$, which has been adopted in SQCD with $N_f=N_c$.  The low-energy degrees of freedom are supplied by the Nambu-Goldstone superfields associated with the breaking of $SU(N_f) \rightarrow SO(N_f)$ instead of magnetic monopoles. Furthermore the flavor-singlet chiral composite ``gauge" superfield is required by the complementarity to balance the $U(1)_R$-anomaly and corresponds to the gauge superfield of $SO(2)$ as the subgroup of $SO(N_c)$.  The low-energy group turns out to be
\begin{equation}
\label{Eq:SummaryFlavorSymBreaking}
SO(N_c)^{loc}\times SU(N_f) \times U(1)_R \rightarrow SO(2)^{loc}\times SO(N_f) \times U(1)_R,
\end{equation}
where the superscript of $loc$ is used to stress the local gauge symmetry.  The $SO(2)^{loc}$ gauge field is the massless composite ``gauge" superfield.

We have extended our discussions of this $SO(N_c)$ theory to other $SO(N_c)$ theories with $N_f\geq N_c-1$ to examine the possibility to use the Nambu-Goldstone superfields as the low-energy degrees of freedom.  In order for the SUSY minimum to be located away from the origin of the moduli space, we have introduced the $Z$-field containing the baryonic degrees of freedom just as used in SQCD with $N_f=N_c$.  The baryonic degrees of freedom we have employed are
\begin{enumerate}
\item color ${\bf N_c}$-plet baryons made of ($N_c-1$)-quarks for $N_f = N_c-1$, and
\item color-singlet baryons made of $N_c$-quarks for $N_f\geq N_c$.
\end{enumerate}
The effective superpotentials take the form of
\begin{equation}\label{Eq:SummaryWeffN}
W_{\rm eff}=S 
\left\{ 
\ln\left[
\frac{
	S^{N_c-N_f-2}\det \left(T\right) f(Z)
}
{
	\Lambda^{3N_c-N_f-6}
} 
\right] 
+N_f-N_c+2\right\},
\end{equation}
equivalent to
\begin{equation}\label{Eq:SummaryW_ADS}
W_{\rm eff}^{\prime}=\epsilon_{N_f-N_c+2}\left( N_f-N_c+2\right) \left[ 
\frac{
	\det \left(T\right) f(Z)
}
{
	\Lambda^{3N_c-N_f-6}
} 
\right]^{1/(N_f-N_c+2)},
\end{equation}
from which $\det (T)f(Z)=0$ is obtained at the SUSY minimum.

With the help of the function of $f(Z)$,  the constraint of $\det (T)f(Z)=0$ replacing the conventional constraint of $\det (T)=0$ allows us to derive $\det (T)\neq 0$ for $f(Z)=0$, where the simplest choice of $f(Z)$ is given by $f(Z)=(1-Z)^\rho$ ($\rho > 0$).  It has been further discussed that the $SO(N_c)$ dynamics with $N_f \geq N_c-1$ can generate
\begin{enumerate}
\item $\det(T)\neq 0$ and $f(Z)=0$ with $f^\prime(Z)=0$ for $N_f = N_c-1$, leading to
\begin{equation}
\label{Eq:SummaryFlavorSymBreakingN-1}
SU(N_f) \times U(1)_R \rightarrow SO(N_f),
\end{equation}
\item $\det(T)\neq 0$ and $f(Z)=0$ with $(f^\prime(Z))^2/f(Z)=0$ for $N_f =N_c$, leading to
\begin{equation}
\label{Eq:SummaryFlavorSymBreakingN}
SU(N_f) \times U(1)_R \rightarrow SO(N_f),
\end{equation}
\item $\det(T)=0$ with ${\rm det}_{N_c}(T) \neq 0$ and $f(Z)=0$ with $(f^\prime(Z))^2/f(Z)=0$ for $N_f \geq N_c+1$, leading to
\begin{equation}
\label{Eq:SummaryFlavorSymBreakingN2}
SU(N_f) \times U(1)_R \rightarrow SO(N_c)\times SU(N_f-N_c) \times U(1)^\prime_R,
\end{equation}
\item $\det(T)=0$ with ${\rm det}_{N_c}(T) \neq 0$ and $f(Z)\neq 0$ for $N_f \geq N_c-1$ with massive $N_f-N_c+2$ quarks.
\end{enumerate}
Since these vacuum configurations are hidden in our effective superpotentials, to single out the configurations, we have used a non-supersymmetric deformation.  Namely, to find $\det(T)\neq 0$ or ${\rm det}_{N_c}(T)\neq 0$ instead of $\det(T)=0$ or ${\rm det}_{N_c}(T)= 0$ in the massless $SO(N_c)$ theories, we have examined how the SUSY broken phase approaches to the SUSY unbroken phase.  As a result, we have shown that the smooth SUSY limit is ensured by $(f^\prime(Z))^2/f(Z)=0$ ($f^\prime(Z)=0$), namely, $\rho > 2$ ($\rho > 1$) for $N_f\geq N_c$ ($N_f=N_c-1$) from the simplest case of $f(Z)=(1-Z)^\rho$ and ${\rm det}(T) \neq 0$ or $\det_{N_c}(T) \neq 0$ is consistently generated in the unbroken SUSY theories.  It is also consistent to have a flavor symmetry breaking in the presence of $Z$, which is not well-defined in the limit of $\det(T)$=$B$=0.  At the symmetry breaking minimum, $Z$ becomes well-defined.  Once these vacuum configurations are known, we simply choose such vacua to analyze our effective superpotential without recourse to the supersymmetric deformation.

Our symmetry breaking respects the complementarity \cite{tHooft,Susskind}, that has helped us find the anomaly-matching property. For large VEV's of squarks instead of vanishing VEV's, this symmetry breaking is understood as a perturbative Higgs phase with squark's VEV's giving masses to quarks and gluinos, describing the same symmetry breaking pattern as Eqs.(\ref{Eq:SummaryFlavorSymBreaking}), (\ref{Eq:SummaryFlavorSymBreakingN-1}), (\ref{Eq:SummaryFlavorSymBreakingN}) and (\ref{Eq:SummaryFlavorSymBreakingN2}).  A Higgs phase (large VEV's) is smoothly connected to a confining phase (VEV's of ${\mathcal{O}}(\Lambda)$) without going through a phase transition. 

It turns out that our superpotentials generally respect
\begin{enumerate}
\item correct vacuum structure with instanton physics in the $SO(N_c)$ theories with $N_f-N_c+2$ massive quarks, where the classical constraint of $f(Z)=0$ is modified to $f(Z)\neq 0$,
\item holomorphic decoupling property utilizing the baryonic degrees of freedom,
\item dynamical breakdown of the flavor $SU(N_f)$ symmetry,
\item consistent anomaly-matching property due to the emergence of the Nambu-Goldstone superfields,
\item consistent decoupling of $S$ by forming a mass term with Tr$(T)$ for $N_f=N_c-2$ and with an appropriate linear combination of Tr$(T)$ and $B$ for $N_f=N_c-1$ and Tr$_{N_c}(T)$ and $B^{[1\cdots N_c]}$ for $N_f\geq N_c$, which is orthogonal to the Nambu-Goldstone mode.
\end{enumerate}
Our proposal that the strongly coupled $SO(N_c)$ theories for $N_f\geq N_c-2$ generate the Nambu-Goldstone superfields is supported by these plausible properties.  It is more plausible if the recent arguments based on the matrix model involving baryonic deformations \cite{MatrixBaryon} utilizing $S$ applied to the present superpotentials would provide the consistent description of the $SO(N_c)$ theories, especially with Eq.(\ref{Eq:SummaryWeffX}) for $N_f=N_c-2$.

\vspace{5mm}
\begin{center}
{\bf ACKNOWLEDGMENTS}
\end{center}
This work is supported by the Grants-in-Aid for Scientific Research on Priority Areas (No 13135219) from the Ministry of Education, Culture, Sports, Science, and Technology, Japan.

\appendix
\section{Instantons}\label{Appendix:InstantonContributions}
The instanton couples to massless quarks and gauginos via their zero-modes.  To obtain the color and flavor singlet form of the corresponding instanton amplitude, the collaboration of two independent instantons are appropriate as in Fig.1, where quarks are converted into scalar quarks by gauge interactions.  This instanton amplitude for gauginos ($\lambda$), $n$ massless quarks ($\psi^{i=1\sim n}$) and $N_f-n$ massive quarks ($\psi^{i=n+1\sim N_f}$) with masses of $m_i$, is defined by the following path-integral:
\begin{eqnarray}
I\left(j, {\bar j}, J, {\bar J}\right)
&=&
\frac{1}{N}
\exp
\left(
-\frac{1}{2}
\int d^4x \sum^n_{i=1}m_i \frac{\delta}{\delta j^i}\frac{\delta}{\delta j^i}
+ {\rm (h.c.)}
\right)
\nonumber \\
&&
\cdot
\int D\psi D{\bar \psi} D\lambda D{\bar \lambda} \exp
\left[
-\int d^4x 
\left(
\sum^{N_f}_{i=1}
\left(
\psi^i
{\not \!\! D}
{\bar \psi}^i
 + \psi^i j^i + {\bar j}^i{\bar \psi}^i
\right)
\right.
\right.
\nonumber \\
&&
\left.
\left.+
\sum^{N_c(N_c-2)/2}_{a=1}
\left(
\lambda^{a}
{\not \!\! D}
{\bar \lambda}^{a}
 + \lambda^a J^a + {\bar J}^a{\bar \lambda}^a
\right)
\right)
\right]
+\cdots,
\label{Eq:PathIntegral}
\end{eqnarray}
in the Euclidean space-time with $g_{\mu\nu}$=diag.($1,1,1,1$), where $N$ is the normalization factor, $j$ ($J$) is the source of $\psi$ ($\lambda$) and ${\not \!\! D}$ = $\sigma^\mu D_\mu$ with $\sigma^0$ =  1, $\sigma^i$ = $i\sigma^{(i)}$ ($i$=1,2,3) for $\sigma^{(i)}$ being the Pauli matrices.  The interaction terms between quarks, gauginos and their superpartners are neglected and the $x$- dependence of the fermions and sources are omitted unless necessary.  After usual computation, we find that
\begin{eqnarray}
I\left(j, {\bar j}, J, {\bar J}\right)
&=&
\frac{1}{N}
{\rm det}_n\left( j_{0i}j_{0j}\right)
\prod^{N_f}_{i=n+1}m_i\int d^4x_i u_{0i}\left( x_i\right) u_{0i}\left( x_i\right)
\prod^{N_c-2}_{\sigma=1}J_{0\sigma}J_{0\sigma}
\nonumber\\
&&\cdot\exp
\left[
\int d^4xd^4y
\left(
{\bar j}\left( x\right) S^\psi_{\bot}\left( x,y\right) j\left( y\right)
+
{\bar J}\left( x\right) S^\lambda_{\bot}\left( x,y\right) J\left( y\right)
\right)
\right]
\nonumber\\
&&
+\cdots,
\label{Eq:PathIntegralResult}
\end{eqnarray}
where $u_{0i}$ is a zero-mode eigen function of the $i$-th quark satisfying $i{\not \!\! D}u_{0i} = 0$, $S^{\psi (\lambda)}_{\bot}$ is the Green's function for $\psi$ ($\lambda$) without zero-mode contributions.  The sources of $j_{0i}$ and $J_{0\sigma}$, respectively, come from zero-modes of $\psi$ and $\lambda$, and are defined as:
\begin{eqnarray}
& & 
j_{0i} = \int d^4x u_{0i}\left( x \right) j\left( x \right),\quad 
J_{0\sigma} = \int d^4x v_{0\sigma}\left( x \right) J\left( x \right),
\label{Eq:ZeroModeSource}
\end{eqnarray}
where $v_{0\sigma}$ is a zero-mode eigen function of the gaugino. In Eq.(\ref{Eq:PathIntegralResult}), the contributions from superpartners of quarks and gauginos are included to cancel $\det({\not \!\! D})$.  We have considered zero-modes from two independent instantons by using two different Grassmann variables, $\xi_{0i}$ and $\xi^\prime_{0i}$, to parametrize $\psi j$ as $\sum^{N_f}_{i=1}(\xi_{0i}j_{0i}+\xi^\prime_{0i}j_{0i})$.

The contributions from zero-modes, $A_0$, are given by
\begin{eqnarray}
&&A_0 
= 
\prod^n_{i=1}
\frac{\delta}{\delta j^i}\frac{\delta}{\delta j^i}
\prod^{N_c-2}_{\sigma=1}
\frac{\delta}{\delta J_\sigma}\frac{\delta}{\delta J_\sigma}
I\left(j, {\bar j}, J, {\bar J}\right)\bigg|_{\rm zero~mode}
\nonumber \\
&&
\hspace{5mm}
= 
\frac{1}{N}
{\rm det}_n\left( u_{0i}u_{0j}\right)
\prod^{N_f}_{i=n+1}m_i\int d^4x_i u_{0i}\left( x_i\right) u_{0i}\left( x_i\right)
\prod^{N_c-2}_{\sigma=1}v_{0\sigma}v_{0\sigma}.
\nonumber \\
&&
\hspace{5mm}
\cdot\exp
\left[
\int d^4xd^4y
\left(
{\bar j}\left( x\right) S^\psi_{\bot}\left( x,y\right) j\left( y\right)
+
{\bar J}\left( x\right) S^\lambda_{\bot}\left( x,y\right) J\left( y\right)
\right)
\right].
\label{Eq:ZeroModeContribution}
\end{eqnarray}
However, for $n\neq 0$, SUSY does not allow this amplitude with $j = {\bar j} = J = {\bar J} = 0$, corresponding to 
\begin{eqnarray}
&&
\langle (\lambda\lambda )^{N_c-2}{\rm det}_n(\psi^i\psi^j)\rangle,
\label{Eq:SUSYBreakingVEV}
\end{eqnarray}
to become constant because Eq.(\ref{Eq:SUSYBreakingVEV}) breaks SUSY if it develops a VEV.  The appropriate amplitude giving a SUSY-preserving constant can be obtained by including the gauge interactions of $\phi^{i\dagger} \lambda \psi^i$ that convert $\lambda\psi^i$ into $\phi^i$. Namely, from the amplitude of
\begin{eqnarray}
\frac{\delta}{\delta j^\dagger_i\left( x\right)}
I\left(j, {\bar j}, J, {\bar J}, j_\phi, j^\dagger_\phi\right),
\label{Eq:ScalarIntegral}
\end{eqnarray}
where $I(j, {\bar j}, J, {\bar J}, j_\phi, j^\dagger_\phi)$ is $I(j, {\bar j}, J, {\bar J})$ with contributionsfrom scalar quarks included and $j_\phi$ and $j^\dagger_\phi$ are the sources of scalar quarks, $\phi$ and $\phi^\dagger$, that enter in 
$I(j, {\bar j}, J, {\bar J}, j_\phi, j^\dagger_\phi )$ as $j^{i\dagger}_\phi\phi^i + \phi^{i\dagger}j^i_\phi$, one can derive that
\begin{eqnarray}
&& \exp
\left(
\sqrt{2}ig\int d^4y \sum^{N_f}_{k=1}
\frac{\delta}{\delta j^k_\phi(y)}
\frac{\delta}{\delta J(y)}
\frac{\delta}{\delta j_k(y)}
\right)
\frac{\delta}{\delta j^{i\dagger}_\phi\left( x\right)}
I\left(j, {\bar j}, J, {\bar J}, j_\phi, j^\dagger_\phi\right)
\nonumber \\
&&
=
\sqrt{2}ig
D\left( x \right)
\frac{\delta}{\delta J(x)}
\frac{\delta}{\delta j^i(x)}
I\left(j, {\bar j}, J, {\bar J}, j_\phi, j^\dagger_\phi\right)
+\cdots,
\label{Eq:ScalarContribution}
\end{eqnarray}
where
\begin{eqnarray}
D\left( x \right) = \int
d^4y
D_F\left( y,x\right)
\label{Eq:FactorD}
\end{eqnarray}
with $D_F(x, y)$ being the Green's function of $\phi$.

It is obvious that, only if $n$=$N_c-2$, the successive use of Eq.(\ref{Eq:ScalarContribution}) can transform the instanton amplitude of Eq.(\ref{Eq:ZeroModeContribution}) to give
\begin{eqnarray}
&&
\langle {\rm det}_{N_c-2}\left(\phi^i\phi^j \right)\rangle
= 
(\sqrt{2}igD)^{2(N_c-2)}A_0\bigg|_{j = {\bar j} = J = {\bar J} = j_\phi = j^\dagger_\phi = 0},
\label{Eq:InstantonScalarVacuum}
\end{eqnarray}
which corresponds to Fig.1.  The amplitude corresponding to a VEV consisting of all scalars like Eq.(\ref{Eq:InstantonScalarVacuum}) can preserve SUSY if it becomes constant.  From Eq.(\ref{Eq:InstantonScalarVacuum}), we finally expect that
\begin{eqnarray}
&&
\prod^{N_f}_{k=N_c-1}\frac{\partial}{\partial m_k}
\langle {\rm det}_{N_c-2}\left(\phi^i\phi^j \right)\rangle
\sim
\Lambda^{2(N_c-2)},
\label{Eq:InstantonVEV}
\end{eqnarray}
leading to
\begin{eqnarray}
&&
\langle {\rm det}_{N_c-2}\left(\phi^i\phi^j \right)\rangle
\sim
\Lambda^{2(N_c-2)}\prod^{N_f}_{i=N_C-1}
\frac{m_i}{\Lambda}.
\label{Eq:InstantonVEVMassive}
\end{eqnarray}
Especially for $N_f=N_c-2$, it is reduced to be
\begin{eqnarray}
&&
\langle {\rm det}\left(\phi^i\phi^j \right)\rangle
\sim
\Lambda^{2N_f}.
\label{Eq:InstantonVEVMassiveNf=Nc-2}
\end{eqnarray}

\section{Broken SUSY}\label{Appendix:BrokenSUSY}
The potential that we have to minimize is given by Eq.(\ref{Eq:VeffN-1}) for $N_f=N_c-1$ and Eq.(\ref{Eq:VeffN}) for $N_f\geq N_c$, which we express as:
\begin{eqnarray}
& & V_{\rm eff}
=
G_T \bigg( 
   \sum_{i=1}^{N_f}  |W_{{\rm eff};i}|^2         
\bigg)
+ 
G_B|W_{{\rm eff};B}|^2
+
G_S |W_{{\rm eff};\lambda} |^2
+
V_{\rm soft}, 
\label{Eq:VeffAll1} \\
& & V_{\rm soft}
=\mu^2\sum_{i=1}^{N_f} \left|\frac{\pi_i}{\Lambda_T}\right|^2
+
\mu^2
\left[
\left| \frac{\pi_B}{\Lambda^{N_B-1}_B}\right|^2
+
\eta\left(
\frac{\pi_B}{\Lambda^{N_B-2}_B}+\frac{\pi^\ast_B}{\Lambda^{N_B-2}_B}
\right)
\right],
\label{Eq:VsoftAll}
\end{eqnarray}
where $N_B=2N_f$ for $N_f=N_c-1$, $N_B=N_c$ and $\eta=0$ for $N_f\geq N_c$.  $W_{{\rm eff};i,B,S}$ are given by
\begin{eqnarray}
\label{Eq:WeffPartial}
&& W_{{\rm eff};i=1\sim N} = \left( 1-\alpha\right)\frac{\pi_S}{\pi_i}, \quad
W_{{\rm eff};i=N+1\sim N_f} =\frac{\pi_S}{\pi_i}, \quad
W_{{\rm eff};B} = n\alpha\frac{\pi_S}{\pi_B},
\nonumber \\
&&
W_{{\rm eff};S} = \ln\left[ \frac{\det(\pi)f(z)}{\pi^{N_f-N_c+2}_S\Lambda^{3N_c-N_f-6}} \right]
\end{eqnarray}
with $N=N_c-1$ for $N_f=N_c-1$ and $N=N_c$ for $N_f\geq N_c$, where $z$ is the VEV of the $Z$ field:
\begin{equation}
z=\frac{\langle 0 \vert B\vert 0 \rangle}{\langle 0 \vert \det(T)\vert 0 \rangle}
=\frac{\pi_B}{\sum^{N_f}_{i=1}\pi_i},
\end{equation}
for $N_f=N_c-1$ and 
\begin{equation}
z=\frac{\langle 0 \vert BT^{N_f-N_c}B\vert 0 \rangle}{\langle 0 \vert \det(T)\vert 0 \rangle}
=\frac{\pi^2_B}{\sum^{N_c}_{i=1}\pi_i},
\end{equation}
for $N_f\geq N_c$, which are expressed as:
\begin{equation}
z=\frac{\pi^n_B}{\sum^N_{i=1}\pi_i},
\label{Eq:z_vs_n}
\end{equation}
with $n=1$ for $N=N_c-1=N_f$ and $n=2$ for $N=N_c\leq N_f$. The coefficients, $\alpha$ and $\beta$, contain the derivative of $f(z)$ defined by
\begin{equation}\label{Eq:alphabeta}
\alpha = z\frac{f^\prime (z)}{f(z)}, \quad \beta = z\alpha^\prime.
\end{equation}

After a little calculus, we find that
\begin{eqnarray}
\label{Eq:VeffZeroPi_a}
&& 
G^\prime_T\vert \pi_i\vert^2\sum^{N_f}_{j=1}\vert W_{{\rm eff};j}\vert^2
+\left( \alpha-1\right)G_TW^\ast_{{\rm eff};a}\frac{\pi_S}{\pi_i}
+\beta G_T\sum^N_{j=1}W^\ast_{{\rm eff};j}\frac{\pi_S}{\pi_j}
\nonumber \\
&&
\qquad+n\beta G_BW^\ast_{{\rm eff};B}\frac{\pi_S}{\pi_B}
+\left( 1-\alpha\right)G_SW^\ast_{{\rm eff};S}
+\pi_i\frac{\partial V_{\rm soft}}{\partial \pi_i}=0,
\end{eqnarray}
from $\partial V_{\rm eff}/\partial\pi_i=0$ with $i=1\sim N$, where the subscript of $i$ is not summed,
\begin{eqnarray}
\label{Eq:VeffZeroPi_i}
&& 
G^\prime_T\vert \pi_i\vert^2\sum^{N_f}_{j=1}\vert W_{{\rm eff};j}\vert^2
-G_TW^\ast_{{\rm eff};i}\frac{\pi_S}{\pi_i}
+G_SW^\ast_{{\rm eff};S}
+\pi_i\frac{\partial V_{\rm soft}}{\partial \pi_i}=0,
\end{eqnarray}
from $\partial V_{\rm eff}/\partial\pi_i=0$ with $i=N+1\sim N_f$ only for $N_f> N_c$, where the subscript of $i$ is not summed,
\begin{eqnarray}
\label{Eq:VeffZeroPi_B}
&& 
G^\prime_B\vert \pi_B\vert^2\vert W_{{\rm eff};B}\vert^2
-n\beta G_T\sum^N_{i=1}W^\ast_{{\rm eff};i}\frac{\pi_S}{\pi_i}
+n\left( -\alpha+n\beta\right) G_BW^\ast_{{\rm eff};B}\frac{\pi_S}{\pi_B}
\nonumber \\
&&
+n\alpha G_SW^\ast_{{\rm eff};S}
+\pi_B\frac{\partial V_{\rm soft}}{\partial \pi_B}=0,
\end{eqnarray}
from $\partial V_{\rm eff}/\partial\pi_B=0$ and
\begin{eqnarray}
\label{Eq:VeffZeroPi_S}
&& 
G^\prime_S\vert \pi_S\vert^2\vert W_{{\rm eff};S}\vert^2
+\left( 1-\alpha\right) G_T\sum^N_{i=1}W^\ast_{{\rm eff};i}\frac{\pi_S}{\pi_i}
+ G_T\sum^{N_f}_{i=N+1}W^\ast_{{\rm eff};i}\frac{\pi_S}{\pi_i}
\nonumber \\
&&
+n\alpha G_BW^\ast_{{\rm eff};B}\frac{\pi_S}{\pi_B}
-\left( N_f-N_c+2\right) G_SW^\ast_{{\rm eff};S}
+\pi_S\frac{\partial V_{\rm soft}}{\partial \pi_S}=0,
\end{eqnarray}
from $\partial V_{\rm eff}/\partial\pi_S=0$. By introducing parameters denoted by $X$ and $Y$:
\begin{eqnarray}
\label{Eq:VeffXY}
&& 
X = G_T\sum^N_{i=1}W^\ast_{{\rm eff};i}\frac{\pi_S}{\pi_i}
-n G_BW^\ast_{{\rm eff};B}\frac{\pi_S}{\pi_B},
\quad
Y = G_T\sum^{N_f}_{i=1}W^\ast_{{\rm eff};i}\frac{\pi_S}{\pi_i},
\end{eqnarray}
and by using Eq.(\ref{Eq:VsoftAll}) for $V_{\rm soft}$ and Eq.(\ref{Eq:WeffPartial}) for $W_{{\rm eff};i,B}$, we finally arrive at
\begin{eqnarray}
&&
G_T\left| \left( 1-\alpha\right)\frac{\pi_S}{\pi_{i=1\sim N}}\right|^2
=
\left( 1-\alpha\right)G_SW^\ast_{{\rm eff};S}
+\beta X
+M^2_T\left| \frac{\pi_i}{\Lambda_T}\right|^2,
\label{Eq:VeffPi_a_xi} \\
&& 
G_T\left| \frac{\pi_S}{\pi_{i=N+1\sim N_f}}\right|
=
G_SW^\ast_{{\rm eff};S}
+M^2_T\left| \frac{\pi_i}{\Lambda_T}\right|^2,
\label{Eq:VeffPi_i_xi} \\
&&
G_B\left| n\alpha\frac{\pi_S}{\pi_B}\right|^2
=
n\left(
\alpha G_SW^\ast_{{\rm eff};S}
-\beta X
\right)
+\mu^2_B\frac{\pi_B}{\Lambda^{N_B-2}_B}
+M^2_B\left| \frac{\pi_B}{\Lambda^{N_B-1}_B}\right|^2,
\label{Eq:VeffPi_B_xi} \\
&& 
\left( N_f-N_c+2\right) G_SW^\ast_{{\rm eff};S}
=Y-\alpha X
+M^2_S\left| \frac{\pi_S}{\Lambda^2_S}\right|^2,
\label{Eq:VeffPi_S_xi}
\end{eqnarray}
where the mass terms of $M^2_{T,B,S}$ and $\mu^2_B$ are defined by
\begin{eqnarray}
\label{Eq:M2}
&& 
M^2_T
=\mu^2+G^\prime_T\Lambda^2_T\sum^{N_f}_{i=1}\vert W_{{\rm eff};i}\vert^2,
\quad
M^2_B
=\mu^2+G^\prime_B\Lambda^2_B\vert W_{{\rm eff};B}\vert^2,
\nonumber \\
&& 
M^2_S
=G^\prime_S\Lambda^2_S\vert W_{{\rm eff};S}\vert^2,
\quad
\mu^2_B = \eta\mu^2.
\end{eqnarray}

It is readily observed that Eqs.(\ref{Eq:VeffPi_a_xi})$\sim$(\ref{Eq:VeffPi_S_xi}) do not permit the solutions with $\pi_{i=N+1\sim N_f}$=$\pi_B$=$\pi_S$=0 for $\pi_{i=1\sim N}\sim \Lambda^2$.  In fact, from Eq.(\ref{Eq:VeffPi_i_xi}), $\pi_S\neq 0$ because of $M^2_T\vert \pi_i\vert\neq 0$ for $i=1\sim N$ and $\pi_{i=N+1\sim N_f,B} = 0$ are not allowed in Eqs.(\ref{Eq:VeffPi_i_xi}) and (\ref{Eq:VeffPi_B_xi}) because of $\pi_S\neq 0$.  Therefore, the solutions indicate the announced property that the flavor-singlet SUSY breaking induces the breakdown of the entire flavor symmetries possessed by the SUSY theory.


\vspace{15mm}
\noindent
\begin{center}
\textbf{TABLE Captions}
\end{center}
\begin{description}
\item{TABLE \ref{Tab:QuntumNumber}:} Quantum numbers of quarks, $Q^i_A$, and gauge fields, $W_{[AB]}$, where {\bf ADJ} stands for the adjoint representation of $SO(N_c)$.
\item{TABLE \ref{Tab:COMP_N-2}:} For $N_f=N_c-2$, quantum numbers of superfields in the Higgs phase and the confining phase, where {\bf SYM} and {\bf ADJ}, respectively, stand for the symmetric and adjoint representations of the $SO$-group, $Q^{\{ij\}}$ denotes the symmetrization with respect to $i$ and $j$ in $Q^i_j$ and $SO(2)$ is the subgroup of $SO(N_c)$ labeled by $A$=1,2.
\item{TABLE \ref{Tab:COMP_N-1}:} The same as in TABLE \ref{Tab:COMP_N-2} but for $N_f=N_c-1$.  The singlet combination in the confining phase is the state orthogonal to the one decoupled with $S$.
\item{TABLE \ref{Tab:COMP_N}:} MThe same as in TABLE \ref{Tab:COMP_N-2} but for $N_f\geq N_c$, where $A,B,a,b$ run from 1 through $N_c$, $i$ runs from $N_c+1$ through $N_f$, Tr$_A(Q)=\sum^{N_c}_{A=1}Q^{\{AA\}}$, Tr$_A(T)=\sum^{N_c}_{A=1}T^{AA}$.  The singlet combination in the confining phase is the state orthogonal to the one decoupled with $S$ and $Q^i_A$ is absent for $N_f=N_c$.
\end{description}

\begin{figure}[!htbp]
\begin{flushleft}
\includegraphics*[1mm,1mm][150mm,70mm]{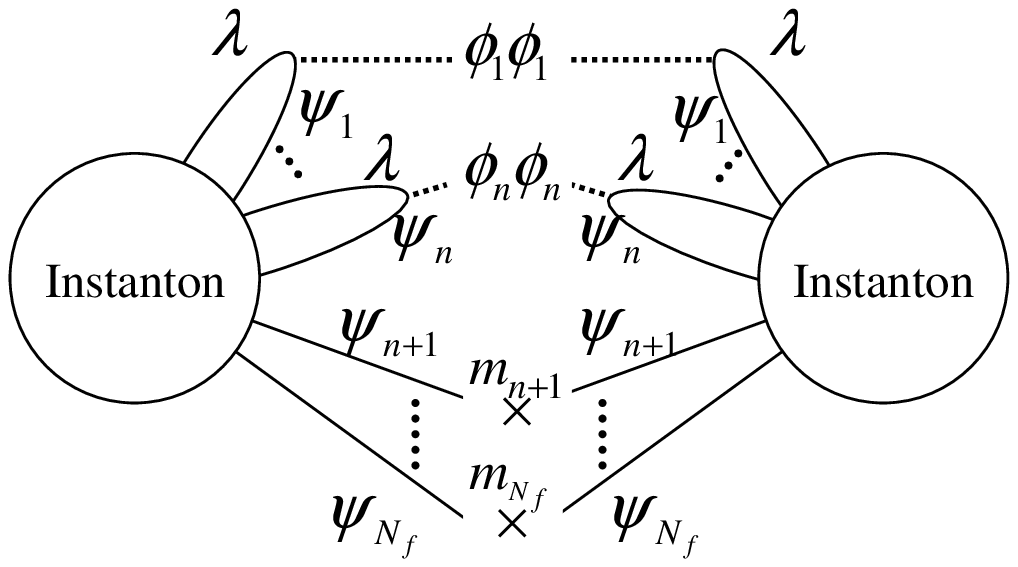}
\end{flushleft}
  \caption{Graphical representation of the instanton amplitude for $N_f\geq N_c-2$, where each instanton couples to $N_c-2$ gauginos ($\lambda$), $n$ massless quarks ($\psi_{i=1\dots n}$) and $N_f-n$ massive quarks ($\psi_{i=n+1\dots N_f}$) and crossed marks indicate the mass-insertion.  This graph is specific to $n=N_c-2$.}
\label{Fig:Instanton}
\end{figure}

\noindent
\begin{center}
\textbf{Figure Caption}
\end{center}
\begin{description}
\item{FIG. \ref{Fig:Instanton}:} Graphical representation of the instanton amplitude for $N_f\geq N_c-2$, where each instanton couples to $N_c-2$ gauginos ($\lambda$), $n$ massless quarks ($\psi_{i=1\dots n}$) and $N_f-n$ massive quarks ($\psi_{i=n+1\dots N_f}$) and crossed marks indicate the mass-insertion.  This graph is specific to $n=N_c-2$.
\end{description}

\noindent
\begin{table}[!htbp]
    \begin{center}
    \begin{tabular}{|c||c|c|c|}
    \hline
        field  & $SO(N_c)$ & $SU(N_f)$ & $U(1)_R$\\ \hline
         $Q^i_A$  & ${\bf N_c}$ & ${\bf N_f}$ & $(N_f-N_c+2)/N_f$ \\
         $W_{[AB]}$ & {\bf ADJ} & {\bf 1} & 1 \\\hline
    \end{tabular}
    \end{center}
    \caption{\small \label{Tab:QuntumNumber} Quantum numbers of quarks, $Q^i_A$, and gauge fields, $W_{[AB]}$, where {\bf ADJ} stands for the adjoint representation of $SO(N_c)$.}
\end{table}
\begin{table}[!htbp]
    \begin{center}
    \begin{tabular}{|c|c||c|c|c|}
    \hline
        Higgs & confining  & $SO(2)$     & $SO(N_f)$   & $U(1)_R$\\ \hline
         $Q^{\{ij\}}$ & $T^{ij}$ & {\bf 1}   & {\bf SYM}  & 0 \\
         $W_{[12]}$   & ${\tilde B}$ & {\bf ADJ}(={\bf 1}) & {\bf 1}   & 1 \\ \hline
    \end{tabular}
    \end{center}
    \caption{\label{Tab:COMP_N-2}\small For $N_f=N_c-2$, quantum numbers of superfields in the Higgs phase and the confining phase, where {\bf SYM} and {\bf ADJ}, respectively, stand for the symmetric and adjoint representations of the $SO$-group, $Q^{\{ij\}}$ denotes the symmetrization with respect to $i$ and $j$ in $Q^i_j$ and $SO(2)$ is the subgroup of $SO(N_c)$ labeled by $A$=1,2.}
\end{table}
\begin{table}[!htbp]
    \begin{center}
    \begin{tabular}{|c|c||c|}
    \hline
        Higgs & confining  & $SO(N_f)$ \\ \hline
         $Q^{\{ij\}}$ &$T^{ij}$ & {\bf SYM} \\
         Tr$(Q)$ & Tr$(T)$, $B$  & {\bf 1} \\ \hline
    \end{tabular}
    \end{center}
    \caption{\label{Tab:COMP_N-1}\small The same as in Table \ref{Tab:COMP_N-2} but for $N_f=N_c-1$.  The singlet combination in the confining phase is the state orthogonal to the one decoupled with $S$.}
\end{table}
\begin{table}[!htbp]
    \begin{center}
    \begin{tabular}{|c|c||c|c|c|}
    \hline
        Higgs & confining  & $SO(N_c)$ & $SU(N_f-N_c)$ & $U(1)_{R^\prime}$\\ \hline
         $Q^{\{AB\}}$ & $T^{ab}$  & ${\bf SYM}$ & ${\bf 1}$ & 0 \\
         $Q^i_A$ &$T^{ia}$ & ${\bf N_c}$ & ${\bf N_f-N_c}$& 1 \\
         Tr$_A(Q)$ & Tr$_A(T),B^{[12\cdots N_c]}$  & ${\bf 1}$ & ${\bf 1}$ & 0 \\ \hline
    \end{tabular}
    \end{center}
    \caption{\label{Tab:COMP_N}\small The same as in Table \ref{Tab:COMP_N-2} but for $N_f\geq N_c$, where $A,B,a,b$ run from 1 through $N_c$, $i$ runs from $N_c+1$ through $N_f$, Tr$_A(Q)=\sum^{N_c}_{A=1}Q^{\{AA\}}$, Tr$_A(T)=\sum^{N_c}_{A=1}T^{AA}$.  The singlet combination in the confining phase is the state orthogonal to the one decoupled with $S$ and $Q^i_A$ is absent for $N_f=N_c$.} 
\end{table}

\noindent

\end{document}